\documentclass[aps,twocolumn,superscriptaddress,showpacs,a4paper]{revtex4}
\usepackage{dcolumn}% Align table columns on decimal point
\usepackage{bm}% bold math
\usepackage{graphicx}% Include figure files
\usepackage{color}
\usepackage{subfigure}
\usepackage{hyperref}
\usepackage{latexsym}
\usepackage{amsthm}
\usepackage{amssymb}
\usepackage{amsmath}
\DeclareGraphicsExtensions{.jpg,.pdf, .mps, .png, .eps, .ps, .EPS,.gif}

\def\bc{\begin{center}} 
\def\ec{\end{center}}
\def\bea{\begin{eqnarray}}
\def\eea{\end{eqnarray}}
\newcommand{\avg}[1]{\langle{#1}\rangle}
\newcommand{\Avg}[1]{\left\langle{#1}\right\rangle}

\begin{document}

\title{ Mesoscopic Structures Reveal the Network Between the  Layers of  Multiplex  Datasets}

\author{Jacopo Iacovacci}
\affiliation{School of Mathematical Sciences, Queen Mary University of London, London, UK}

\author{Zhihao Wu}
\affiliation{School of Computer and Information Technology, Beijing Jiaotong University, Beijing, China}

\author{Ginestra Bianconi}
\affiliation{School of Mathematical Sciences, Queen Mary University of London, London, UK}

\begin{abstract}
Multiplex networks describe a large variety of complex systems, whose elements (nodes) can be connected by different types of interactions forming different layers (networks) of the multiplex. Multiplex networks include social networks, transportation networks or biological networks in the cell or in the brain. Extracting relevant information from these networks is of crucial importance for solving challenging inference problems and for characterizing the multiplex networks microscopic and mesoscopic structure. 
Here we propose an information theory method to extract the network between the layers of multiplex datasets, forming a  ``network of networks". We build an indicator function, based on the entropy of network ensembles, to characterize the mesoscopic similarities between the layers of a multiplex network and we use clustering techniques to characterize the communities present in this network of networks. We apply the proposed method to study the Multiplex Collaboration Network formed by scientists collaborating on different subjects and publishing in the  Americal Physical Society (APS) journals. The analysis of this dataset reveals the interplay between the collaboration networks and the organization of knowledge in physics. 
\end{abstract}

\pacs{89.75.Fb, 89.75.Hc and 89.75,-k}

\maketitle
\section{Introduction}

Multiplex networks \cite{boccaletti2014structure,kivela2014multilayer} describe a large number of complex systems where the interactions are of different nature. They are formed by a set of $N$ nodes interacting through $M$ different layers (networks). 
Recently, multiplex networks have been used to  characterize a large variety of systems, including social networks \cite{szell2010multirelational}, transportation network \cite{cardillo2012modeling}, collaboration networks \cite{menichetti2014weighted,nicosia2014measuring}, and brain networks \cite{bullmore2009complex}.
Extracting relevant information from multiplex networks is central for characterizing their microscopic and mesoscopic structure   \cite{mucha2010community,battiston2014structural,de2015structural}, for solving challenging inference problems, and for devising good centrality measures \cite{halu2013multiplex,sola2013eigenvector,de2015ranking}.

Structural correlations  are ubiquitous in multilayer networks and can be a powerful tool to extract information from them. 
For example the overlap of the links \cite{bianconi2013statistical} in the different layers of multiplex networks has been observed in systems as different as in-silico societies \cite{szell2010multirelational}, multilayer airport networks \cite{cardillo2012modeling} or citation-collaboration networks \cite{menichetti2014weighted}. Moreover it was recently shown \cite{menichetti2014weighted} that using the information on the link overlap it is possible to extract information that cannot be extracted if the single layers are taken in isolation.
Other examples of correlations encoded in multiplex network structures include correlation between the degrees of the same node in different layers \cite{min2014network}, and the  activity distribution of the nodes \cite{nicosia2014measuring,cellai2015}. 

All these structural correlations reflect local properties of multiplex networks. Nevertheless, in complex networks, significant information is encoded in their mesoscale structure, i.e. their organization into several clusters or communities \cite{fortunato2010community,bianconi2014triadic}.

Recently new modularity measures for multilayer networks \cite{mucha2010community} have been proposed and new multiplex community detection algorithms have been formulated  \cite{de2014identifying} based on methods devised for single networks \cite{fortunato2010community}. Alternatively inference methods have been proposed to decompose a single network in different layers with distinct community structure \cite{valles2014multilayer} or to visualize multiplex networks \cite{de2014multilayer}. 
Moreover it has been recently observed that the communities on different layers of a multiplex networks typically overlap among each others, forming mesoscale structures that span across different layers.
This phenomenon is central for generalizing the concept of community to multilayer networks \cite{mucha2010community,de2014identifying} and modeling the emergence of communities \cite{battiston2015emergence}.\\ 
In this paper our aim is to characterize the correlations of multiplex networks at the mesoscopic scale, and to use this information in order to build a network between the layers of multiplex datasets. In particular we propose an information theory measure $\widetilde{\Theta}^{S}$, able to define similarities between the layers of a multiplex respect to their mesoscopic structures. This similarity is more significant when groups of nodes densely connected with each others are simultaneously present on different layers, forming overlapping communities.  
This measure is based on the concept of network entropy \cite{bianconi2008entropy,bianconi2009entropy,peixoto2012entropy} and extends the $\Theta$ measure presented in \cite{bianconi2009assessing}. Using the similarity $\widetilde{\Theta}^{S}$, here we propose a method  for  extracting  the network between the layers of multiplex networks. We apply the proposed method to the characterization of the  American Physical Society (APS) Collaboration Multiplex Networks extracted from the APS dataset \cite{dataset}. The scientific collaboration networks have been studied extensively in the context of single networks \cite{redner98,newman2001PNAS,newman2001PRE, arenas2004,lee2010}.
Nevertheless, additional relevant information can be extracted if they are analyzed as a multilayer structure \cite{menichetti2014weighted,nicosia2014measuring,de2015ranking}. The Collaboration Multiplex Networks are formed by the authors of the APS papers, and by layers corresponding to the Physics and Astronomy Classification Scheme (PACS) codes \cite{pacs}. In particular two authors are linked on layer $\alpha$ if they have co-authored a paper with PACS code corresponding to layer $\alpha$. Since the PACS codes are organized in  hierarchical levels we constructed two APS Collaboration Multiplex Networks corresponding to layers describing either the first or the second level of the PACS hierarchy. 
The analysis performed on the APS Collaboration Multiplex Networks has allowed us to characterized the network between the layers of these multiplex networks, and to investigate the same dataset at different levels of resolution with respect to the number of layers.\\
The paper is structured as follows: in Section II we define the indicator measure $\widetilde{\Theta}^S$; in Section III we test the measure on two different multiplex benchmark models of two-layer network with communities; in Section IV we use our measure to analyze the community structure of the APS Collaboration Multiplex Network at two hierarchical levels of the PACS code; { in Section V we compare the results obtained with $\widetilde{\Theta}^S$ with results obtained using other similarity measures on the same dataset}; finally in Section VI we give the conclusions.

\section{Definition of $\widetilde\Theta^{S}$}

Our goal here is to construct an information theory indicator function $\tilde{\Theta}^S$ to characterize the   similarity in the mesoscopic structure  of the layers of a multiplex network. 
This indicator function is based on the entropy of network ensembles \cite{bianconi2008entropy,bianconi2009entropy,peixoto2012entropy,bianconi2009assessing}, a quantity which plays a key role when inference problems are addressed using an unbiased information theory approach \cite{peixoto2012entropy,bianconi2009assessing}.
In this section we define how the indicator function  $\widetilde\Theta^{S}$ is defined.
We consider a multiplex network formed by $N$ nodes $i=1,2\ldots, N$ and $M$ layers $\alpha=1,2,\ldots,M$. The structure of the multiplex network is characterized by $M$ adjacency matrices ${\bf a}^{\alpha}$ of elements $a_{ij}^{\alpha}=1$ if node $i$ is connected to node $j$ in layer $\alpha$, or $a_{ij}^{\alpha}=0$ otherwise.
We indicate with  $k_{i}^{\alpha}$ the degree of a node $i$ on layer $\alpha$, i.e. the number of neighbors that node $i$ has on $\alpha$.
The nodes having degree $k_i^{\alpha}=0$ in layer $\alpha$, are the isolated nodes, i.e. nodes that are not connected to any other node in the layer $\alpha$, also called \cite{nicosia2014measuring} in the context of multilayer networks ``inactive" nodes in layer $\alpha$. Conversely all the nodes with $k_i^{\alpha}>0$ are called ``active" nodes in layer $\alpha$.

We assume that each node $i$ of layer $\alpha$ has a  characteristic $q_{i}^{\alpha} \in \{1,\ldots, Q^{\alpha}\}$. The quantity $q_i^{\alpha}$ can for example indicate the community to which the node $i$ belongs. 
More in general $q_i^{\alpha}$ can represent any feature of the nodes in layer $\alpha$. 
Starting from this information we can classify the nodes in $P^{\alpha}$ classes $p_i^{\alpha}\in \{1,\ldots, P^{\alpha}\}$ which take into account at the same time the information about the degree of the nodes and their characteristic $q_i^{\alpha}$. 
This is the minimal assumption to capture the structure of networks with communities induced by the characteristics $q^{\alpha}=\{q_i^{\alpha}\}_{i=1,2\ldots,N}$, and strong heterogeneities in the degree. Considering only the partition induced by the characteristics would imply that in the network we do not consider the structure induced by the degrees, which is clearly not a viable option for broadly distributed networks. 

Including other features of the nodes to define node classes could be a viable option. In this case  the characteristics $q^{\alpha}$ will take into account different features which  might depend on the specific network under consideration. Therefore here we take the class $p_i^{\alpha}$ to be a function of degree $k_i^{\alpha}$ and of the characteristic $q_i^{\alpha}$, i.e. $p_i^{\alpha}=f(k_i^{\alpha},q_i^{\alpha})$. 
   The block structure of the network  induced by the classes $p_{i}^{\alpha}=f(k_i^{\alpha},q_i^{\alpha})$  is described by the matrices $\textbf{e}^{\alpha}$ of elements $e^{\alpha}(p,p')$ indicating the total number of links on the layer $\alpha$ between nodes of class $p$ and nodes of class $p'$. We define the entropy $\Sigma_{k^{\alpha},q^{\alpha}}$ \cite{bianconi2008entropy,bianconi2009entropy,peixoto2012entropy,bianconi2009assessing} of a layer $\alpha$ as the logarithm of the number of graphs preserving the block structure $\textbf{e}^{\alpha}$ in a given layer.
By considering the number of graphs preserving a given block structure, we have that this entropy takes the simple expression,

\bea
\Sigma_{k^{\alpha},q^{\alpha}}=\log\left[\prod_{p<p'}\left(\begin{array}{c} n_p^{\alpha}n_{p'}^{\alpha} \label{entropy}\\ e^{\alpha}(p,p')\end{array}\right)\prod_p \left(
\begin{array}{c} n_p^{\alpha}(n_{p}^{\alpha}-1)/2 \\ e^{\alpha}(p,p)\end{array}\right)\right],
\eea
where 
\bea
\hspace*{-5mm}e^{\alpha}(p,p')&=&\sum_{i,j} a_{ij}^{\alpha}\delta\left[p_i^{\alpha}(k_i^{\alpha},q_i^{\alpha}),p\right]
\delta\left[p_j^{\alpha}(k_i^{\alpha},q_i^{\alpha}),p'\right],
\eea   
for $p\neq p'$, and $e(p,p),n(p)$ given respectively by 
\bea
\hspace*{-5mm}e^{\alpha}(p,p)&=&\sum_{i<j} a_{ij}^{\alpha}\delta\left[p_i^{\alpha}(k_i^{\alpha},q_i^{\alpha}),p\right]
\delta\left[p_j^{\alpha}(k_i^{\alpha},q_i^{\alpha}),p\right],
\eea
and 
\bea
n_p^{\alpha}&=&\sum_i \delta\left[p_i^{\alpha}(k_i^{\alpha},q_i^{\alpha}),p\right],
\eea
with $\delta[x,y]$ indicating the Kronecker delta.
     The entropy  $\Sigma_{k^{\alpha},q^{\alpha}}$ is a measure to assess how much information is encoded in the constraint imposed to the network i.e. the block structure $\textbf{e}^{\alpha}$. The smaller is the entropy the smaller is the number of networks that share the block structure $\textbf{e}^{\alpha}$. Therefore the smaller is the entropy of an ensemble the larger is the level of information encoded by the constraint. 
If for a given assignment of the characteristics $\{q^{\alpha}_i\}$ the entropy is much smaller than in a random hypothesis (when the characteristics are reshuffled randomly between the nodes), then the network structure reflects the characteristic assignment $\{q^{\alpha}_i\}$ and thus the characteristics $\{q^{\alpha}_i\}$  capture relevant information respect to the network structure.
Following this argument the quantity $\Theta$ proposed in \cite{bianconi2009assessing}, which is based on the entropy of network ensembles, has been shown to be an unbiased indicator able to quantify the specificity of a generic layer $\alpha$ to the assignment $q_i^{\alpha}$. This information theory quantity is defined as:  

\begin{equation}
\Theta_{k^{\alpha},q^{\alpha}}=\frac{E_{\pi}[\Sigma_{k^{\alpha},\pi(q^{\alpha})}]-\Sigma_{k^{\alpha},q^{\alpha}}}{\sqrt{E_{\pi}[(\Sigma_{k^{\alpha},\pi(q^{\alpha})}-E_{\pi}[\Sigma_{k^{\alpha},\pi(q^{\alpha})}])^{2}]}},
\end{equation}
where $E_{\pi}[...]$ is the expected value over random uniform permutations $\pi(q^{\alpha})$ of the node characteristics $q^{\alpha}$ in layer $\alpha$.

Here we propose to use this quantity to compare the similarity between the different layers in a multiplex network. Indeed we can consider the characteristics $q^{\beta}$ of the nodes in layer $\beta$ as an induced feature of nodes in layer $\alpha$ and measure by the corresponding indicator $\Theta_{k^{\alpha},q^{\beta}}$ how much information the characteristics $q^{\beta}$ contain respect to the node structure of layer $\alpha$. 
In particular the indicator $\Theta_{k^{\alpha},q^{\beta}}$ is given by  

\begin{equation}
\Theta_{k^{\alpha},q^{\beta}}=\frac{E_{\pi}[\Sigma_{k^{\alpha},\pi(q^{\beta})}]-\Sigma_{k^{\alpha},q^{\beta}}}{\sqrt{E_{\pi}[(\Sigma_{k^{\alpha},\pi(q^{\beta})}-E_{\pi}[\Sigma_{k^{\alpha},\pi(q^{\beta})}])^{2}]}}.
\end{equation}

Therefore $\Theta_{k^{\alpha},q^{\beta}}$ measures the specificity of the layer $\alpha$ respect to the particular set $q^{\beta}$, which is the assignment of the characteristics of the nodes on layer $\beta$.

When one considers a single layer, the entropy is independent on the choice adopted for classifying isolated (inactive) nodes in layers belonging to multiplex networks.
In fact, we can either group all the isolated nodes in a single class or each isolated node in a different class, and the entropy value given by Eq. $(\ref{entropy})$ does not change because the isolated nodes have no links attached to them. 
Instead the indicator function $\Theta_{k^{\alpha},q^{\alpha}}$ might depend on this choice because its construction involves several reshuffling of the characteristics of the nodes.
 
When comparing different layers of a multiplex network, the nodes that are active in one layer might not be active in another layer. Nevertheless, the information carried by the activity of the node might be significant. For example if two layers have very different activity patterns, it might occur that   the nodes inactive in one layer form a well defined cluster in the other layer resulting in a very  significant information that is important to capture.
Therefore to distinguish between nodes active and inactive in a layer it is a very convenient choice to classify all the inactive nodes in one layer under a given common characteristic.
A similar type of argument can be made about connected clusters of small sizes, which are ``quasi-isolated" as the nodes belonging to connected clusters of size 2 or 3 etc.
Depending on the number of such clusters it might be convenient to classify also nodes in connected components of size 2 or 3 etc. into given common characteristics as we will show in the next sections using the concrete examples of the APS Collaboration Multiplex Networks.
Here, if not stated otherwise, we will consider  the case in which the features $q^{\alpha}$ indicates the community  of the nodes in layer $\alpha$ and the characteristic $p^{\alpha}_i$ takes a different value for each distinct pair $(k_i^{\alpha},q_i^{\gamma})$ where $k_i^{\alpha}\neq 0$, while all the nodes with $k_i^{\alpha}=0$ form another class of nodes. 
%We focus here in the case in which the feature $q^{\alpha}=\{q_i^{\alpha}\}_{i=1,2\ldots,N}$ indicates the community of the nodes in layer $\alpha$ and the class of  $p^{\alpha}_i$ of node $i$ in layer $\alpha$ takes a different value for each distinct pair $(k_i^{\alpha},q_i^{\gamma})$ as long as the node $i$ is not isolated $k_i^{\alpha}>0$, and it  belongs  to a community of  more than two nodes.  All the isolated nodes belong a the same class $\tilde{p}$. All the nodes belonging to a two-node community belong to another class $\hat{p}$. This choice is dictated by the need to reduce the influence of isolated nodes and of the two-node clusters.
%We focus here in particular in the case in which the characteristic $q^{\alpha}_i$ takes into account  the size of the connected cluster to which node $i$ belongs, and the community assignment of the node $i$. Moreover the classes $p_i^{\alpha}$ take a  different value for each distinct pair $(k_i^{\alpha},q_i^{\gamma})$.

In order to compare the level of information carried in layer $\alpha$ by the community structure in layer $\beta$, $q^{\beta}$,  with the level of information carried by the proper community structure, $q^{\alpha}$, we define the quantity
\begin{equation}
\widetilde{\Theta}_{\alpha,\beta}= \frac{\Theta_{k^{\alpha},q^{\beta}}}{\Theta_{k^{\alpha},q^{\alpha}}}.
\end{equation}

 \begin{figure}[h!]
\includegraphics[width= 8 cm]{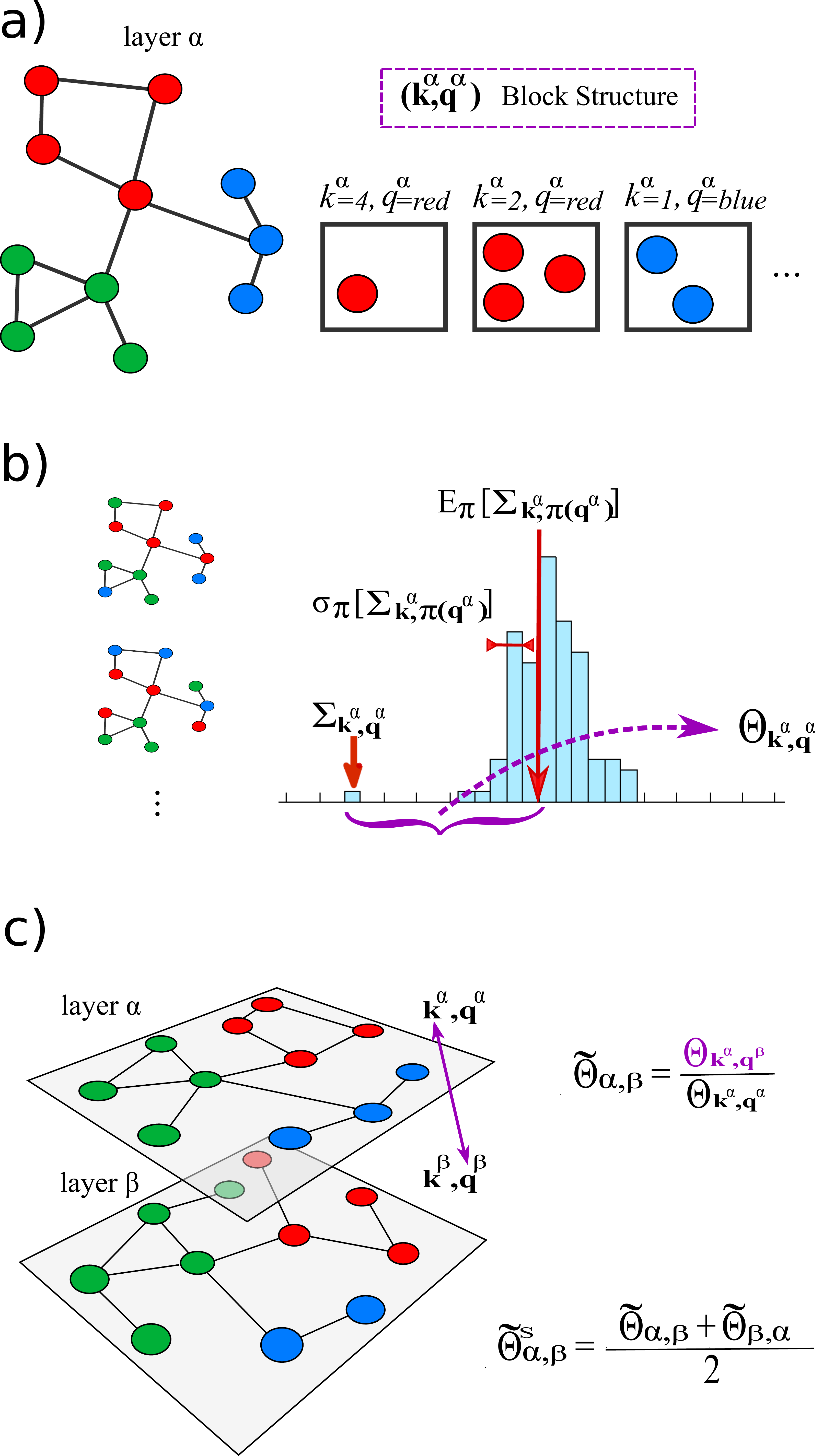} 
  \caption{(Color online) Diagram showing the method. Panel a): We consider a layer $\alpha$ in a multiplex network and we define the node classes  $p^{\alpha}=(k^{\alpha},q^{\alpha})$, where $k^{\alpha}$ indicates the node degrees  
  and $q^{\alpha}$ the node characteristics on the layer $\alpha$. {These classes induce a block structure in the network specified  by  the number of links between the nodes of each class and the number of links connecting the nodes in different classes}. Panel b):  The entropy $\Sigma_{k^{\alpha},q^{\alpha}}$ given by Eq. $(\ref{entropy})$ is calculated and compared with the entropy distribution obtained in a random hypothesis, by performing random uniform permutations $\pi(q^{\alpha})$ of the characteristics $q^{\alpha}$ of the nodes and subsequently measuring the $\Sigma_{k^{\alpha},\pi(q^{\alpha})}$ values. The mean $ E_{\pi}\left[\Sigma_{k^{\alpha},\pi(q^{\alpha})}\right]$ and standard deviation $\sigma_{\pi}\left[\Sigma_{k^{\alpha},\pi(q^{\alpha})}\right]$ of the entropy distribution is thus calculated. The indicator function $\Theta_{k^{\alpha},q^{\alpha}}$ measures the difference between $\Sigma_{k^{\alpha},q^{\alpha}}$ and $E_{\pi}\left[\Sigma_{k^{\alpha},\pi(q^{\alpha})}\right]$ in units of $\sigma_{\pi}\left[\Sigma_{k^{\alpha},\pi(q^{\alpha})}\right]$. Panel c): Given a second layer $\beta$, $\widetilde\Theta_{\alpha,\beta}$ characterizes the information about the structure in layer $\alpha$, carried by the characteristics of nodes in layer $\beta$. In order to define a symmetric indicator function of the similarity between the layers $\alpha$ and $\beta$ we define the indicator $\widetilde\Theta^S_{\alpha,\beta}$ that symmetrizes the indicator function  $\widetilde\Theta_{\alpha,\beta}$. }
  \label{fig:method}
\end{figure}

The quantity $\widetilde{\Theta}_{\alpha,\beta}$ is a measure of how layer $\beta$ is similar to $\alpha$ respect to the community assignment $\textbf{q}$. If $\widetilde{\Theta}_{\alpha,\beta}=1$ the community structure $q^{\beta}$, proper of layer $\beta$, carries the same level of information for the structure of layer $\alpha$ as the community structure $q^{\alpha}$, proper of the layer $\alpha$. It is important to notice that the matrix $\widetilde{\Theta}$ in principle is not symmetric. We can construct the symmetric measure $\widetilde{\Theta}^{S}_{\alpha,\beta}$ by symmetrizing the quantity $\widetilde{\Theta}_{\alpha,\beta}$ i.e. by defining

\begin{equation} 
\widetilde{\Theta}_{\alpha,\beta}^{S}=\frac{\widetilde{\Theta}_{\alpha,\beta}+\widetilde{\Theta}_{\beta,\alpha}}{2}.
\end{equation}

This is a symmetric measure indicating how similar layer $\alpha$ and layer $\beta$ are with respect to their community structure.
In Figure $\ref{fig:method}$ we give a  schematic summary of the method used to construct the similarity measure $\widetilde{\Theta}_{\alpha,\beta}^{S}$.

In a given multiplex network, we can then analyze the entire symmetric matrix $\widetilde{\Theta}^S$ measuring the similarity between the community structure of the layers.
This matrix  characterizes the entire multiplex network at the layer level, reducing the information about the network structures to one matrix of similarity between the layers.

In the following Section we will first test this measure on multiplex network benchmark models with non trivial community structure, then in the subsequent Section we will focus on characterizing the APS Collaboration Multiplex Networks where the layers are the collaborations networks  of scientists using different PACS numbers.

In this paper we are mostly concerned about similarities in the community structure of the layers of a  multiplex network,  nevertheless it has to be stressed that the proposed approach and similarity measure $\widetilde{\Theta}_{\alpha,\beta}^{S}$ is general and it can be used by considering any available feature of the nodes related to the structure of the layers.

\section{Testing $\widetilde\Theta^{S}$ on Benchmark Models}

In order to validate on a well defined multiplex architecture our similarity measure $\widetilde\Theta^{S}$ respect to the community structures of different layers of a multiplex network, we have developed two benchmark models with communities.
In particular we want to construct benchmark multiplex network models with a controlled level of overlap between the communities in different layers. Given in a generic multilayer the community assignment $q^{\alpha}$ of the nodes on each layer $\alpha$, 
we define the community overlap as 
\bea
O_{c}=\frac{2}{M(M-1)}\frac{1}{N}\max_{\{\pi\}}\left\{\sum_{\alpha<\beta}^{M}\sum_{i=1}^{N} \delta\left[q_i^{\alpha},\pi(q_i^{\beta})\right]\right\},
\eea 
where $M$ indicates the total number of layers and $N$ indicates the total number of nodes, $\delta[x,y]$ indicates the Kronecker delta and the maximum is taken over all the permutations $\pi(q^{\beta})$ of the label of the communities in layer $\beta$. 

\begin{figure}
\includegraphics[width=\columnwidth]{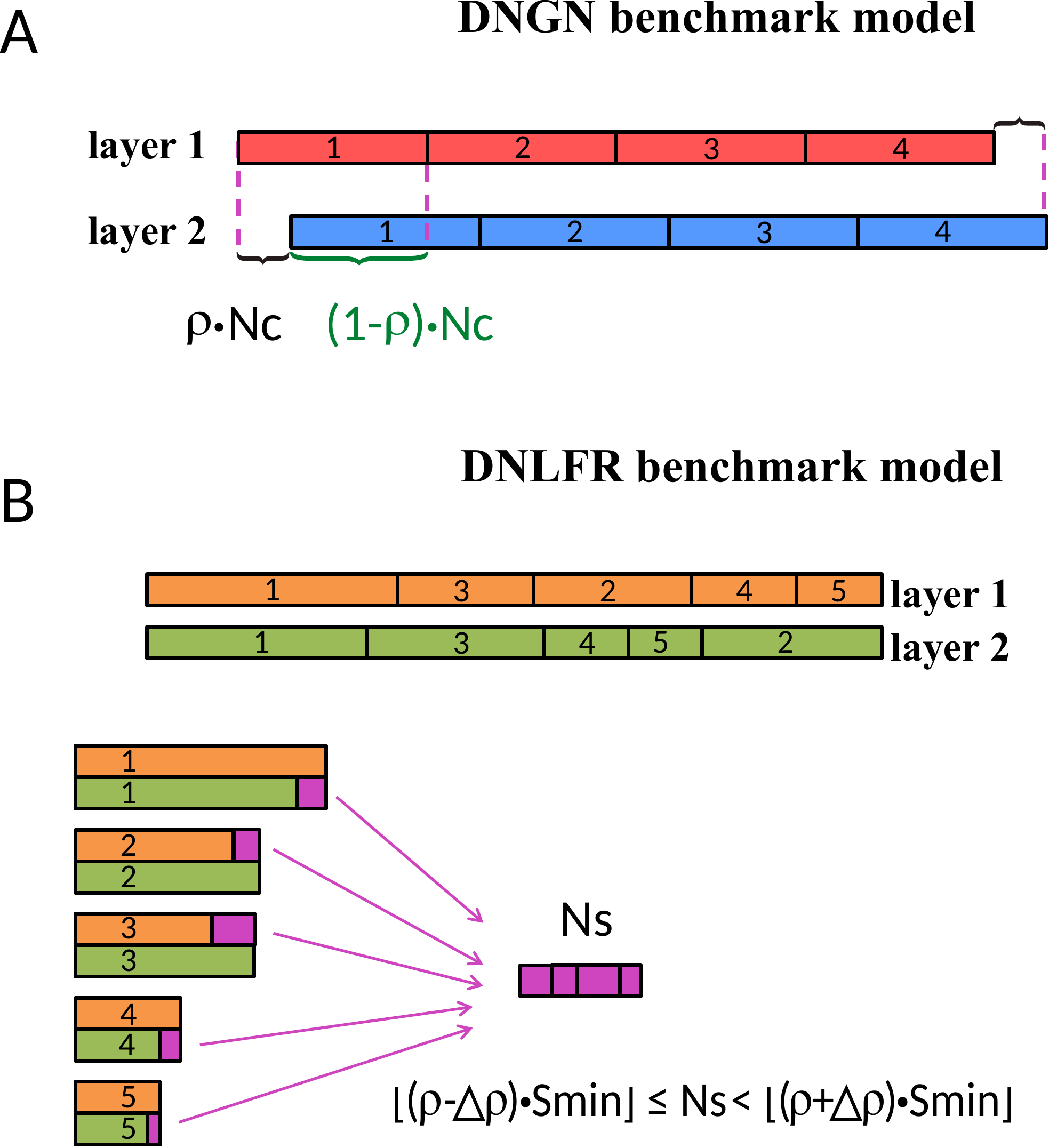} 
  \caption{(Color online) Schematic of the benchmark models DNGN and DNFLR. Panel (A). The DNGN benchmark model: nodes on both layers (blue and red) are divided into four communities of equal size $N_{c}$, labelled from 1 to 4. Each  community of layer 1 overlaps for a fraction of $(1-\rho)\cdot N_{c}$ nodes with its corresponding community in layer 2. Panel (B). The DNLFR benchmark model: on each layer  $Q=5$ non homogeneous communities are generated and labelled from 1 to 5 according to their size (left). For a given $\rho$ the total number of nodes which do not overlap between communities of the same label, $N_{s}$, has values $\lfloor(\rho-\Delta \rho)\cdot S_{min}\rfloor\leq N_{s}<\lfloor(\rho + \Delta \rho)\cdot S_{min}\rfloor$, where $\lfloor...\rfloor$ is the floor function and $S_{min}$ is the minimum bound of the power-law distribution from which the community sizes in the two layers are extracted.}
  \label{fig1.1}
\end{figure}

We define two benchmark models (see Figure \ref{fig1.1}) based respectively on the Girvan and Newman (GN) \cite{girvan2002community}  model and on the Lancichinetti~-~Forunato~-~Radicchi (LFR) model \cite{lancichinetti2008benchmark}, which are very well established benchmarks for single networks with communities. The proposed benchmarks are designed to tune the overlap of communities between different layers of simple  multiplex networks having respectively homogeneous or heterogeneous degree distribution and community size distribution.

For the first benchmark model, the Duplex Network GN model (DNGN) we construct a duplex network (a multiplex network made of two layers) in which each layer is formed by a GN network realization. Therefore each of the layers is formed by $N$ nodes divided into $4$ equal size clusters of size $N_c$. 

 The network in each layer is a random network in which each node has a  probability $p_{in}$ to link to nodes of its same community  and a probability $p_{out}$ to link to nodes outside its community. In particular  we have chosen $p_{in}$ and $p_{out}$  in order to have for each node, a  mean  degree $\avg{k}=16$ and a mean number of links outside the community given by $\avg{k_{out}}=4$. The layers generated in this way have a well defined community structure and they are essentially random respect to other network characteristics. 
The characteristic $q_i^{\alpha}$ indicates the community to which a node $i$ belongs on layer $\alpha=1,2$.
 Here we consider the possible correlations  existing between the community assignment $q_i^{[1]}$ and $q_i^{[2]}$ in the two layers. This community assignment allows us to tune in a control way the level of overlap between the communities.
In particular we label the nodes $i=1\ldots, N$ in layer 1 according to the following community assignment  $q_i^{\alpha}$,
\bea
q_i^{[1]}=\left \lceil\frac{i}{N_c}\right\rceil.
\eea 
where the brackets $\lceil x\rceil$ in the right end side of this expression indicate the ceiling function of  $x$.
Therefore we have, for $N=128$ and $N_c=32$,
\bea
q_i^{1}=\left\{\begin{array}{lcl}1 &\mbox{for}& i\in [1,32]\nonumber \\
2& \mbox{for} & i\in[33, 64] \nonumber \\
3 &\mbox{for}& i \in [65,96] \nonumber \\
4 & \mbox{for} &i\in [97,128] \end{array}.\right.
\eea
The community assignment in layer 2 will not be in general the same of layer 1.
In order to model overlap of communities we perform a simple ``shift'' of the labels, parametrized with the parameter $\rho>0$. In particular we take 
\bea
q_i^{[2]}=\left\{\begin{array}{ccc}\left\lceil\frac{i-\rho N_c}{N_c}\right \rceil&\mbox{if}& \left\lceil\frac{i-\rho N_c}{N_c}\right \rceil>0\nonumber \\
&&\nonumber \\
\frac{N}{N_c}&\mbox{if} &\left\lceil\frac{i-\rho N_c}{N_c}\right \rceil=0\end{array}.\right.
\eea

In general the control parameter $\rho$ takes values $0\leq \rho \leq 0.5$. If $\rho = 0$ there is no ``shift'' between the layer partitions (they perfectly match); if $\rho > 0$  each community in the first layer overlaps with the corresponding one in the second layer for a fraction of nodes equal to $(1-\rho)\cdot N_{c}$; thus $\rho \cdot N_{c}$ is the number of ``shifted'' nodes per community. When $\rho=0.5$, $N=128$ and $N_c=32$, we have
\bea
q_i^{2}=\left\{\begin{array}{lcl}1 &\mbox{for}& i\in [17,48]\nonumber \\
2& \mbox{for} & i\in[49,80] \nonumber \\
3 &\mbox{for}& i \in [81,112] \nonumber \\
4 & \mbox{for} &i\in [1,16]\cup[113,128] \end{array}. \right.
\eea
Therefore $\rho=0.5$ describes the maximum ``shift'' between the community of the two layers: each community in the first layer shares $16$ nodes with its corresponding community in the second layer.
Given a value of $\rho$ the overall community overlap in the network can be easily calculated, being $O_{c}=(1-\rho)$, and in the case of maximum ``shift'' we obtain $O_{c}=0.5$.

For the second benchmark model the Duplex Network LFR model (DNLFR), we have taken a duplex network in which the single layers are constructed according to the LFR model \cite{lancichinetti2008benchmark}.
\begin{enumerate}
\item
The network in the first layer is a LFR network, formed by $Q$ communities. The communities are labelled according to their size in descending order.\\
\item
The network in the second layer is a LFR network with $Q$ communities generated using the same parameters used for the network in the first layer. Additionally we require that the network in the second layer satisfies a further condition, which allows us to modulate the overlap between the communities in the two layers.
Specifically, for each second layer candidate, we first label the communities according to their size in descending order. Then we compare each of them to the corresponding one in the first layer (panel B Figure \ref{fig1.1}). We calculate the number of ``shifted'' nodes $N_s$ given by the sum of the absolute values of the difference between the corresponding communities sizes, i.e.
\bea
N_s=\sum_{l=1}^Q \left|n^{[1]}_l-n^{[2]}_l\right|,
\eea
where $n_l^{\alpha}$ is the size of the community $l$ in layer $\alpha$.
Finally we retain the candidate network as the second layer of the duplex network only if 
\bea
\lfloor(\rho-\Delta \rho)\cdot S_{min}\rfloor\leq N_{s}<\lfloor(\rho + \Delta \rho)\cdot S_{min}\rfloor,
\label{s}
\eea where $\lfloor\ldots \rfloor$ is the floor function. Here   $\rho$ and $\Delta \rho$ are control parameters of the benchmark model that modulate the overlap of the communities, and $S_{min}$ in Eq. $(\ref{s})$ is the parameter that in the LFR model fixes the lower bound of the community sizes.
In this way if one considers a sufficient number of multiple realizations of the multilayer, and a sufficiently low value of $\Delta \rho$, one gets 
\bea
\avg{N_{s}} \simeq \lfloor \rho \cdot S_{min}\rfloor.
\eea
\item
Finally, the nodes  are relabelled in both layers in order to allow the
 maximum community overlap.  In particular the labels are reassigned in such a way that the   common number  of nodes in the communities that have the same label in the two layers, is equal to the minimum of the two community  sizes. (see Figure $\ref{fig1.1}$.)\\   
Therefore the average community overlap of the benchmark network is dependent on $\rho$ and, for a significant number of realizations and low enough values of $\Delta \rho $, is given by  
\bea
\Avg{O_c}=1-\frac{\Avg{N_{s}}}{N}\simeq 1-\frac{ \lfloor \rho \cdot S_{min}\rfloor}{N}.
\eea
 \end{enumerate}
 
\begin{figure}
\includegraphics[width=\columnwidth]{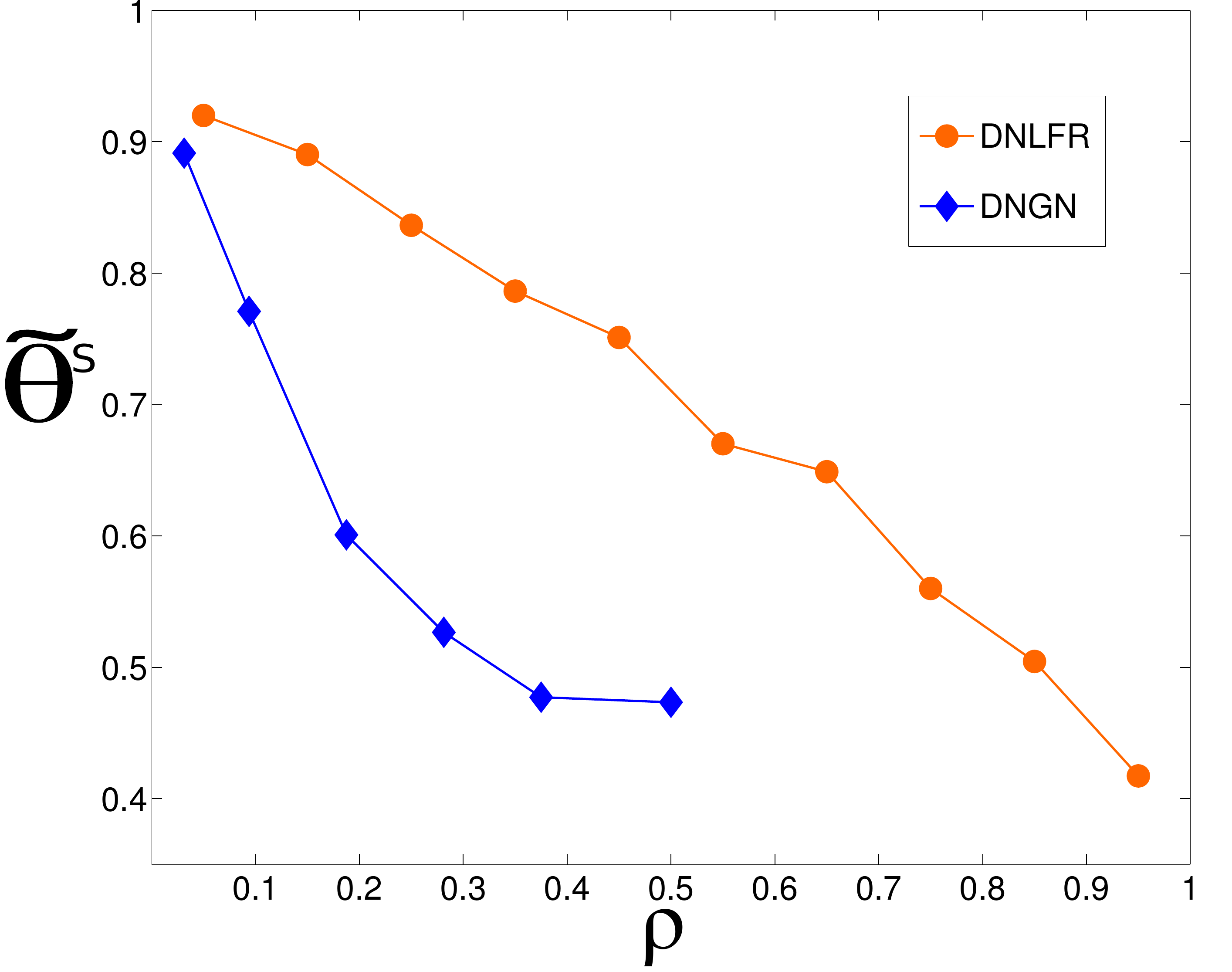} 
  \caption{(Color online) The similarity measure $\widetilde\Theta^{S}$ between the two layers of the DNGN (blue diamonds) and DNFLR (orange circles) benchmark models is measured as a function of the control parameter $\rho$. When $\rho$ increases the total community overlap between the layers decreases and $\widetilde\Theta^{S}$ decreases monotonically both in the case of homogeneous-size communities (DNGN) and in the case of heterogeneous-size communities (DNFLR). Each data point is averaged over $50$ benchmark realizations. For the DNFLR model the parameter $\Delta \rho$ was set to 0.05.}
  \label{fig1}
\end{figure}

 In order to test the performance of the similarity measure $\widetilde\Theta^{S}$, we apply this measure to the two duplex network benchmarks, for different values of $\rho$.
 Since $\rho$ modulates the level of community overlap between the layers we expect that the similarity measure $\widetilde\Theta^{S}$ is larger for lower value of $\rho$ (corresponding to larger community overlap $O_c$ between the layers) and smaller for larger values of $\rho$ (corresponding to smaller community overlap $O_c$ between the layers).
In Figure $\ref{fig1}$ we show the dependence  $\widetilde\Theta^{S}$ as a function of $\rho$ for the two proposed benchmark models. In both cases the displayed values $\widetilde\Theta^{S}$ are averaged over $50$ benchmark realizations.\\
For the DNGN benchmark, we considered $N=128$, $N_c=32$ and $\rho \leq 0.5$. The similarity measure $\widetilde\Theta^{S}$ is monotonically decreasing with $\rho$. For the DNLFR benchmark the two single layers are generated according to the LFR algorithm with parameters $N = 600$ (number of nodes) and $Q=5$ (number of communities). The size of each community is taken from a power-law distribution with lower bound $S_{min} = 60$, upper bound $S_{max} = 180$, and power-law exponent $\tau_{1} = 1.5$. inside the communities the node degree distribution is also extracted from a power-law distribution with parameters $k_{max} = 50$ (maximum degree), $\tau_{2} = 2.6$ (power-law exponent), $\avg{k}= 16$ (average degree). For building the DNLFR network we used $\Delta\rho = 0.05$ and $\rho\leq 0.95$. Also in the case of the DNLFR benchmark, where the size of the communities is heterogeneous, $\widetilde\Theta^{S}$ decreases monotonically with $\rho$.

This result shows that in benchmark models in which the community overlap is modulated by an external control parameter, $\widetilde\Theta^{S}$ decreases  together with the community overlap. 
Since in general measuring the community overlap involves an optimization over a permutation of the community assignment, measuring the community overlap can be very costly numerically.
In this situation calculating $\widetilde\Theta^{S}$ could instead give an alternative way to assess the similarity between the layers of a multiplex network.

In Section V, using the concrete examples of the APS Collaboration Multiplex Networks, we will compare the similarity measure  $\widetilde\Theta^{S}$ to other existing measures introduced to compare different community assignments in single layers.

\section{The network between the layers of the APS Collaboration Multiplex Networks}
\label{APS}
In this Section, we use the similarity matrix $\widetilde\Theta^{S}$ to analyze the APS Collaboration Multiplex Networks.
These multiplex networks are extracted from the APS collaboration dataset \cite{dataset} recording all the bibliometric information about the papers published in the APS journals.

The network is formed by a set of $N$ nodes representing the APS authors. 
Since there is no agreement on disambiguation techniques for the author names, we have identified each author with the initials of his/her first name and last name.
 The layers correspond to different Physics and Astronomy Classification Scheme (PACS) codes \cite{pacs} describing the subject of the papers.
Two authors are linked in a given layer $\alpha$ if they are co-authors of at least one paper having the PACS number corresponding to layer $\alpha$. 
Since PACS numbers are organized in a hierarchical way (the first digit of the number indicates the general field of physics while the second digit specifies the ambit inside that field), we have constructed two multiplex networks whose layers correspond respectively to the first and second hierarchical level of the PACS codes.  
The APS Collaboration Multiplex Network related to the first level of the hierarchy of PACS codes is made of $M_{1} =10$ layers each one describing the collaboration network in a general field of physics. 
The APS Collaboration Multiplex Network at the second level of the hierarchy is made of $M_{2} =66$ layers each one describing the collaboration network in a specific ambit of physics (second level of the PACS code hierarchy).

In extracting the APS Collaboration Multiplex Networks we considered all the papers until 2014 with less than ten co-authors. This threshold was introduced to exclude papers coming from  big collaborations  that follow different statistical properties with respect to the rest of the dataset. With this threshold, our dataset includes a consistent fraction of the whole dataset ($\simeq 97\%$ of the total number of papers) and a number of authors $N = 180,539$.

The layers of the APS Collaboration Multiplex Networks  are characterized by a significantly different activity pattern of the nodes. Moreover roughly 0.7\% of nodes belong to connected components of size 2 while only about 0.006\% of the nodes belongs to connected components of size 3. Therefore we consider here the case in which the characteristics $\{q_i^{\alpha}\}$ indicate the community of the nodes in layer $\alpha$ and the class $p^{\alpha}_i$ of node $i$ in layer $\alpha$ takes a different value for each distinct pair $(k_i^{\alpha},q_i^{\gamma})$ as long as the node $i$ is not isolated $k_i^{\alpha}>0$, and it  belongs  to a community of  more than two nodes.  All the isolated nodes belong a the same class $\tilde{p}$. All the nodes belonging to a two-node community belong to another class $\hat{p}$.

\begin{figure}[h!]
\includegraphics[width=\columnwidth]{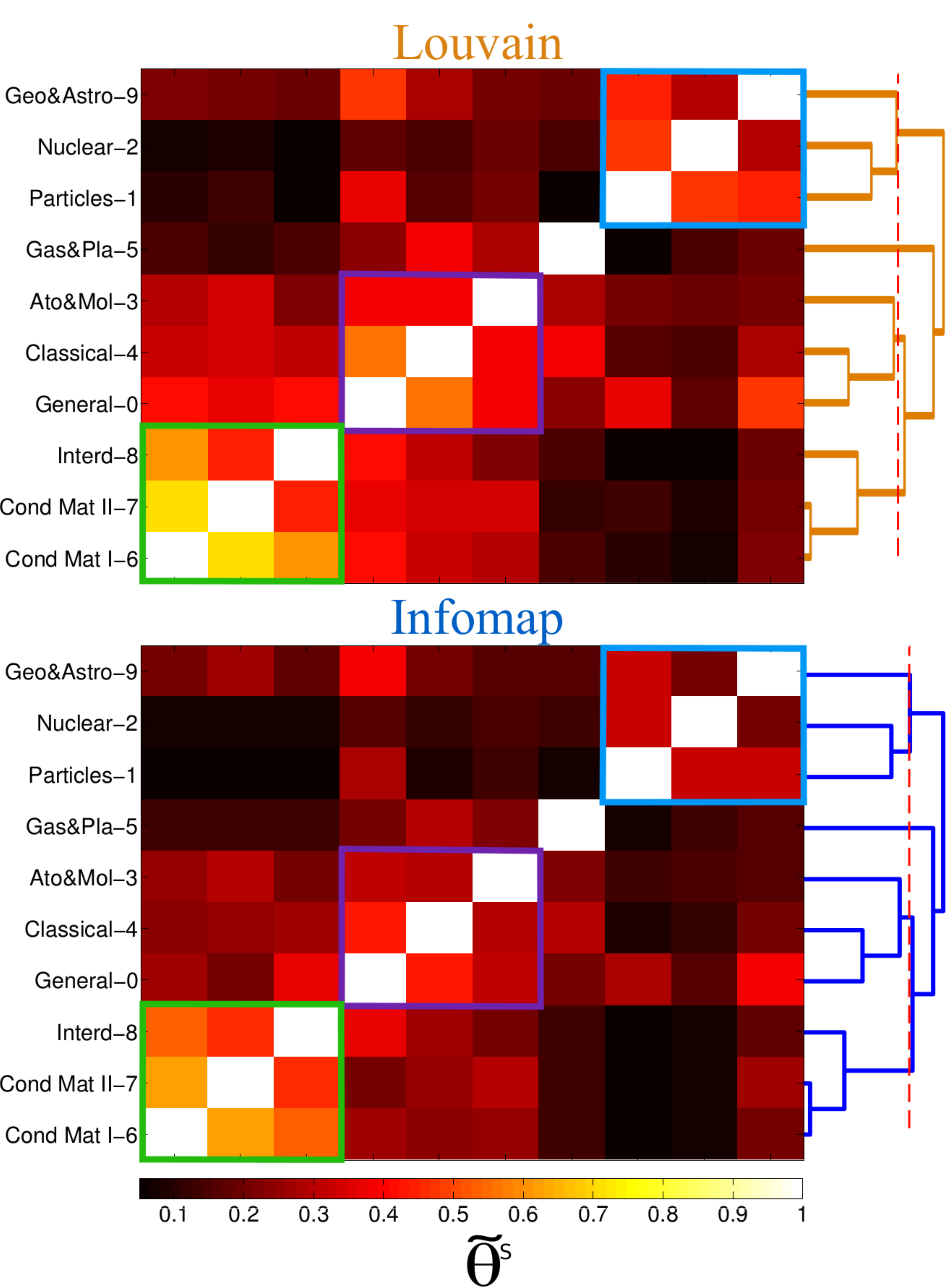} 
  \caption{(Color online) The similarity matrices of elements $\widetilde\Theta^{S}_{\alpha,\beta}$ calculated respectively using the Louvain and the Infomap community detection algorithms are plotted for the APS Collaboration Multiplex Network with the $M_1=10$ layers indicating the collaboration network at the first level of the PACS hierarchy. Each layer refers to a general field of Physics (see  Table $\ref{t1}$ for the legend of the layer acronyms).
  The dendrogram between the layers is shown on the left of each matrix $\widetilde\Theta^{S}_{\alpha,\beta}$. The dashed line on top of the dendrogram indicates the partition that correspond to the optimal value of the weighted modularity given by Eq. $(\ref{Q})$.}
  \label{fig3}
\end{figure}

Let us first characterize the mesoscale similarities between the $M_1=10$ layers of the APS Collaboration Multiplex Network  in the main subjects of physics, described by the first level of the PACS code hierarchy.
The similarity matrix $\widetilde\Theta^{S}$ is constructed in two different ways, using either the Informap community detection algorithm \cite{rosvall2007information} and the Louvain algorithm \cite{blondel2008fast}, and averaging in both cases over $350$ random permutations of the community assignments. For simplicity we will refer to these two matrices as Infomap-$\widetilde\Theta^{S}$ and Louvain-$\widetilde\Theta^{S}$. The two matrices are reported in Figure \ref{fig3} in the form of heat-maps. The patterns shown by the two heat-maps are very similar, denoting that from a qualitatively point of view the measure $\widetilde\Theta^{S}$ is not affected by the choice of the algorithm used to perform the community detection for the network under study. We can observe that, in general, clusters in the APS Collaboration Multiplex Network extend across multiple layers. As expected, layers describing collaborations in general or interdisciplinary fields such as General Physics or Interdisciplinary Physics, which often involve people from different specific ambits of physics, show high values of $\widetilde\Theta^{S}$ respect to several other layers while more specific fields, such as Gases$\&$Plasma, show lower values of $\widetilde\Theta^{S}$ respect to the other layers. 

Given this similarity measure between the layers of the multiplex, one can build a network of networks whose nodes represent the $M_1=10$ networks of collaboration in general fields of physics and whose weighted edges are the values $\widetilde\Theta^{S}_{\alpha,\beta}$ and represent the similarity between the $M_1$ networks respect to their community structure.
This network of layers is thus a weighted fully-connected network showing itself a significant community structure and revealing how the pattern of collaboration between scientists is organized across different fields of physics.
In order to characterize this community structure between the layers of the multiplex network, we perform a hierarchical clustering analysis starting from the dissimilarity matrix $d$ of elements $d_{\alpha,\beta}$ given by 
\begin{equation}
d_{\alpha,\beta}=1-\left|\widetilde\Theta^{S}_{\alpha,\beta}\right|.
\label{d}
\end{equation}
Specifically, we use the average linkage clustering method which gave the best cophenetic correlation coefficient compared to other clustering methods\cite{clust1,clust2,clust3}. According to the average method the distance $d_{c}(C_1,C_2)$ between two clusters $C_1$ and $C_2$ is defined as the average distance between all pairs of layers in the two clusters:   
\begin{equation}
d_{c}(C_1,C_2)=\frac{1}{{\cal N}(C_1){\cal N}(C_2)}\sum_{\alpha \in C_1}\sum_{\beta\in C_2}d_{\alpha,\beta}
\end{equation}
where ${\cal N}(C_i)$ indicates the number of layers in cluster $C_i$.

\begin{table}[h!]
\begin{tabular}{|l|l|l|}
\hline
\multicolumn{1}{|c|}{\textbf{Acronym}}&\multicolumn{1}{c|}{\textbf{PACS}} & \multicolumn{1}{c|}{\textbf{Field}}                                                                                                     \\ \hline
General-0  &00                              & \begin{tabular}[c]{@{}l@{}}General\end{tabular}                                                   
\\ \hline
Particles-1                    &10             & \begin{tabular}[c]{@{}l@{}}Physics of Elementary \\ Particles and Fields\end{tabular}                                                    \\ \hline
    Nuclear-2               &20             & \begin{tabular}[c]{@{}l@{}}Nuclear Physics\end{tabular}                                                   \\ \hline
           Ato\&Mol-3    &30                   & \begin{tabular}[c]{@{}l@{}}Atomic and Molecular\\ Physics\end{tabular}                                                                   \\ \hline
Classical-4&40                                  & \begin{tabular}[c]{@{}l@{}}Electromagnetism, Optics,\\ Acoustic, Heat Transfer,\\ Classical Mechanics and \\ Fluid Dynamics\end{tabular} \\ \hline
Gas\&Pla-5 & 50                               & \begin{tabular}[c]{@{}l@{}}Physics of Gases, Plasmas\\ and Electric Discharges\end{tabular}                                              \\ \hline
Cond Mat I-6 &60                                & \begin{tabular}[c]{@{}l@{}}Condensed Matter: \\Structural, \\ Mechanical and Thermal\\ Properties\end{tabular}                              \\ \hline
Cond Mat II-7   &70                          & \begin{tabular}[c]{@{}l@{}}Condensed Matter: \\Electronic Structure,\\ Electrical, \\Magnetic and Optical properties\end{tabular}             \\ \hline
Interd-8           &80                     & \begin{tabular}[c]{@{}l@{}}Interdisciplinary Physics and Related \\ Areas of Science and Technology\end{tabular}                         \\ \hline
       Geo\&Astro-9   &90                        & \begin{tabular}[c]{@{}l@{}}Geophysics, \\
       Astronomy and Astrophysics                                                                                                 \end{tabular}  \\ \hline
\end{tabular}
\caption{The acronyms used in this study for the PACS number at the first level of the PACS hierarchy, the corresponding PACS numbers and corresponding general fields of Physics.}
\label{t1}
\end{table}
In Figure $\ref{fig3}$, together with  the matrices Infomap-$\widetilde\Theta^{S}$ and Louvain-$\widetilde\Theta^{S}$ we show the dendrograms resulting from the hierarchical clustering analysis of the respective dissimilarity matrices Infomap-$d$ and Louvain-$d$. 
In order to define an optimal partition of the layers into communities, we looked for the agglomerative stage of the cluster hierarchy at which the weighted modularity $Q$ \cite{newman2006modularity} is maximized, $Q$ defined as:
\begin{equation}
Q=\frac{1}{\langle \eta \rangle M}\sum_{\alpha \neq \beta}^{M}\left(\left|\widetilde\Theta^{S}_{\alpha,\beta}\right|-\frac{\eta_{\alpha}\eta_{\beta}}{\langle \eta \rangle M}\right)\delta \left[\sigma_{\alpha}\sigma_{\beta}\right],
\label{Q}
\end{equation}
where $\sigma_{\alpha}$ labels the community in which layer $\alpha$ is, $\delta[x,y]$ indicates the Kronecker delta and $\eta_{\alpha}$, $\langle \eta\rangle$ are given respectively by 
\begin{eqnarray}
\eta_{\alpha}&=&\sum_{\beta\neq\alpha}\left|\widetilde\Theta^{S}_{\alpha, \beta}\right|,\nonumber \\
\langle \eta \rangle&=&\frac{1}{M}\sum_{\alpha} \eta_{\alpha}.
\end{eqnarray}

 As shown in Figure \ref{fig3} the optimal partition found is the same either when using the Infomap algorithm or the Louvain algorithm to perform the community detection in the layers of the multiplex. The analysis reveals that the first layers clustering together are Condensed Matter I$\&$II and Interdisciplinary Physics and they form the first block (green coloured box); the second block includes General Physics, Classical Physics, Atomic and Molecular Physics (purple coloured box); in the third block Particles Physics, Nuclear Physics and Geophysics$\&$Astrophysics group together (cyan coloured box). The layer related to Gases$\&$Plasma Physics is isolated and can be considered as a block by itself.

\begin{figure}[h!]
\includegraphics[width=\columnwidth]{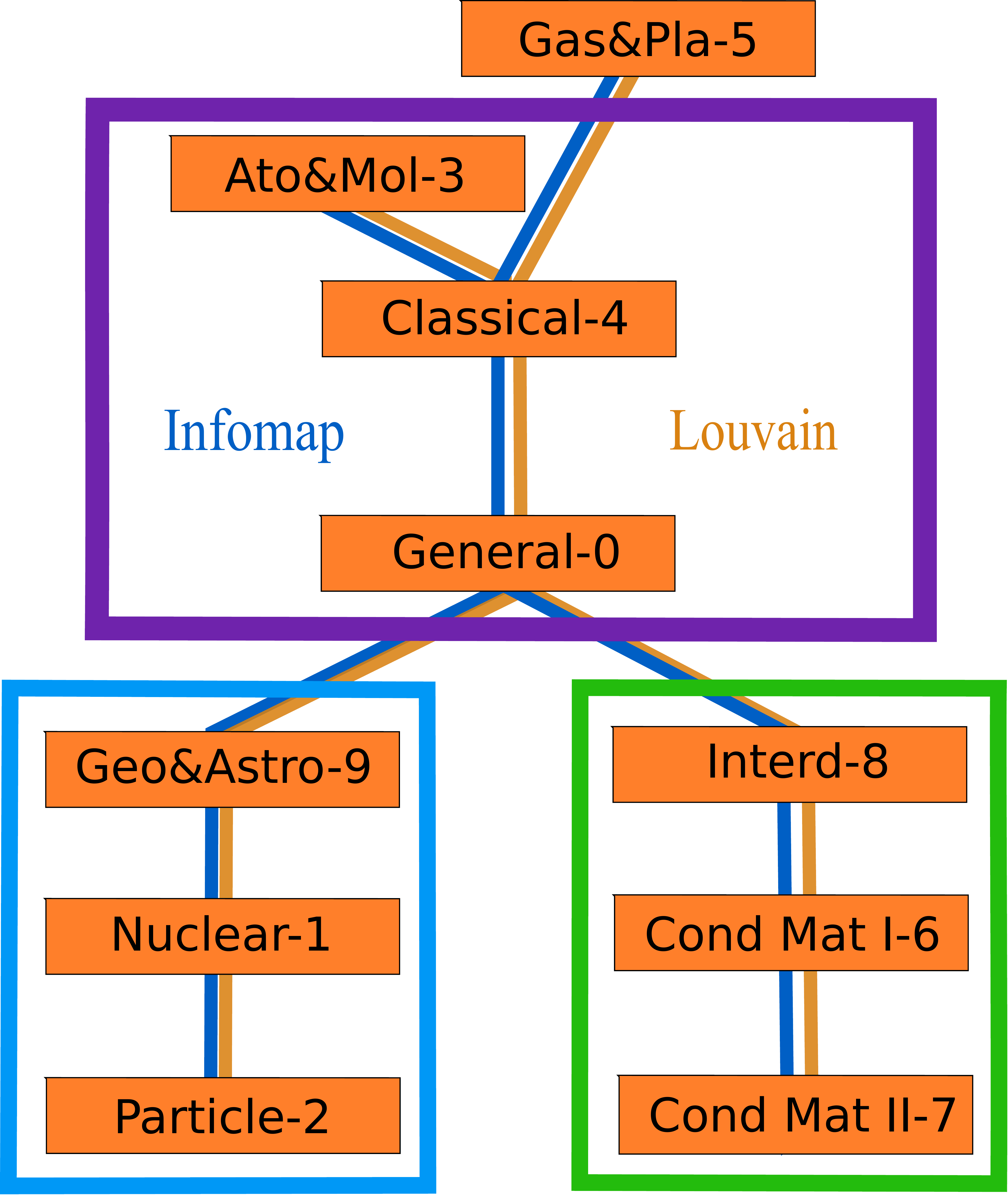} 
  \caption{(Color online) Minimal Spanning Tree (MST) using the dissimilarity measure $d$ in the case of Infomap-$d$ dissimilarity (blue) and in the case of Louvain-$d$ dissimilarity (ocher). The block structure obtained with the hierarchical clustering analysis is also showed. 
  %The community structure of the layer General Physics spans both on the Astrophysics\&Geophysics layer and on the Interdisciplinary Physics layer, and results in the inter-connections between the three main blocks.
  }
  \label{fig4}
\end{figure}

Once revealed the block (community) structure an interesting issue is to characterize the Minimal Spanning Tree (MST) that allows us to identify the layers which connect the blocks together. Therefore we construct the  MST using the dissimilarity measure $d$ defined in Eq.~$(\ref{d})$ calculated either using  the Infomap or the Louvain clustering algorithm. The two MSTs are identical (Figure \ref{fig4}) and this  confirm the robustness of the results with respect to the community detection algorithm used.  We can see that the collaboration layer of General Physics connects the three main blocks  together.\\

\begin{table}[h]
\begin{tabular}{lllll}
\cline{1-4}
\multicolumn{1}{|c|}{\textbf{Block 1}}                                                & \multicolumn{1}{c|}{\textbf{Block 2}}                                                               & \multicolumn{1}{c|}{\textbf{Block 3}}                                                                          & \multicolumn{1}{c|}{\textbf{Block 4}}                                                                &  \\ \cline{1-4}
\multicolumn{1}{|l|}{\begin{tabular}[c]{@{}l@{}}\\ Cond Mat I-6\\Cond Mat II-7\\Interd-8\\ \end{tabular}} & \multicolumn{1}{l|}{\begin{tabular}[c]{@{}l@{}}\\General-0\\Ato\&Mol-3\\ Classical-4\end{tabular}} & \multicolumn{1}{l|}{\begin{tabular}[c]{@{}l@{}}\\Particles-1\\Nuclear-2\\Geo\&Astro-9 \end{tabular}} & \multicolumn{1}{l|}{\begin{tabular}[c]{@{}l@{}}Gas\&Pla-5 \end{tabular}} &  \\ \cline{1-4}
                                                                                      &                                                                                                     &                                                                                                                &                                                                                                      &  \\
                                                                                      &                                                                                                     &                                                                                                                &                                                                                                      & 
\end{tabular}
\caption{Clusters between the $M_1=10$ layers of the APS multiplex network corresponding to the first level of the PACS hierarchy (see for the legend of the layer acronym Table $\ref{t1}$). The clusters have been obtained from the dendrograms shown in Figure $\ref{fig3}$, cut in order to obtain the partition that optimizes the weighted modularity $Q$ defined in Eq. (\ref{Q}).}
\label{t2}
\end{table}

\begin{figure*}
\includegraphics[width=\columnwidth]{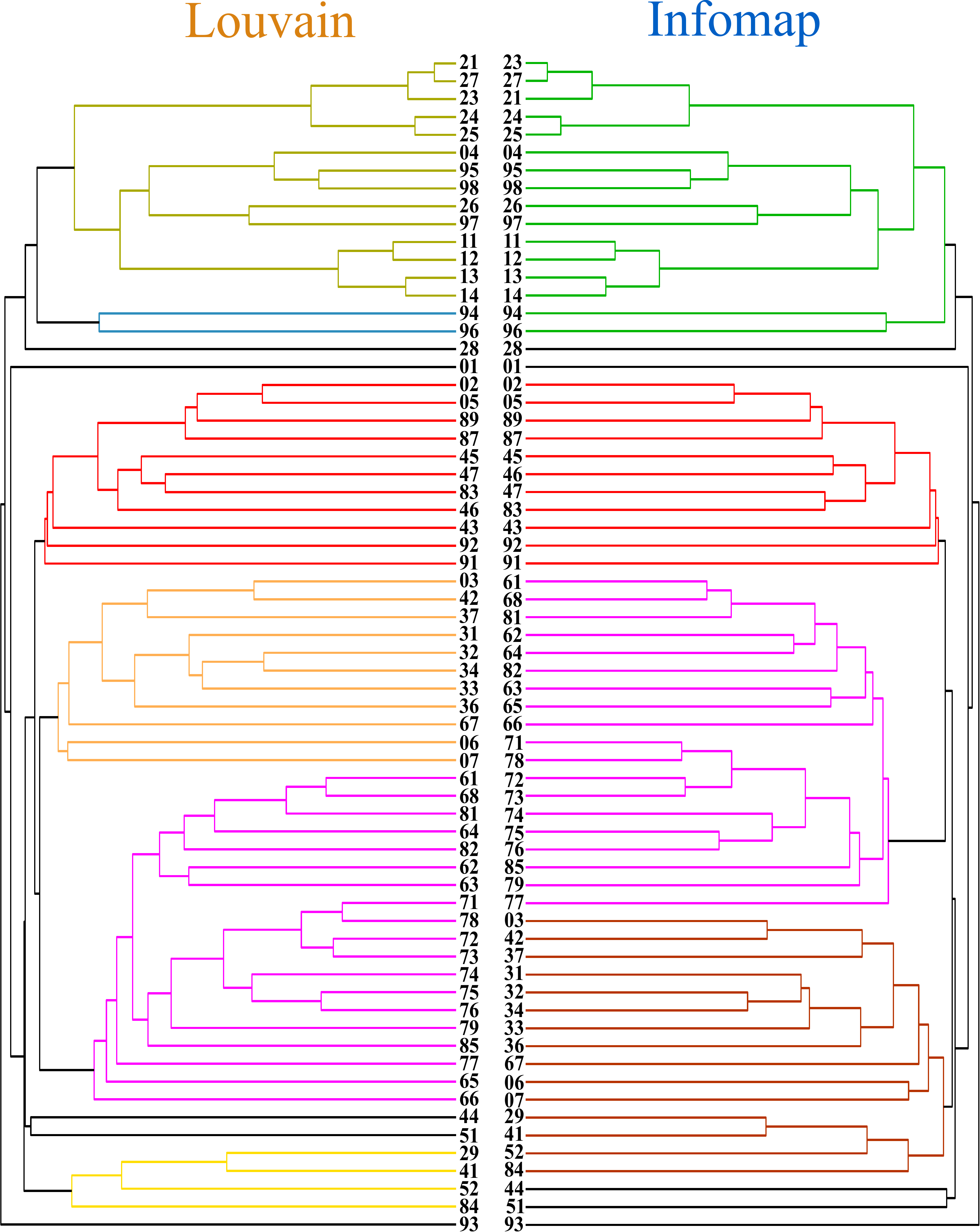} 
  \caption{(Color online) Hierarchical clustering of the APS Collaboration Multiplex Network in which each layer represents a collaboration network in a specific area of physics, as described by the second hierarchical level of the PACS code. We show the two dendrograms obtained respectively from the Louvain-$\widetilde\Theta^{S}_{\alpha,\beta}$ (left) and from the Infomap-$\widetilde\Theta^{S}_{\alpha,\beta}$ (right). In each dendrogram the communities found at the optimal partition (maximum of $Q$) are represented as branches of the same colors.     
}
  \label{fig5}
\end{figure*}

In order to have a deeper understanding of the results previously found we now consider the multiplex network of scientific collaborations where the layers are related to the PACS code at the second level of the PACS hierarchy. For this multiplex network we have calculated the similarity matrix $\widetilde\Theta^{S}$ between the $M_2=66$ layers and found the optimal partition into communities according to the score function $Q$, following an analogous procedure to the one used previously for first level of the PACS hierarchy. To calculate $\widetilde\Theta^{S}_{\alpha,\beta}$ we have performed  averages over $350$ random permutations of the community assignments.
\begin{figure*}
\includegraphics[width=16cm]{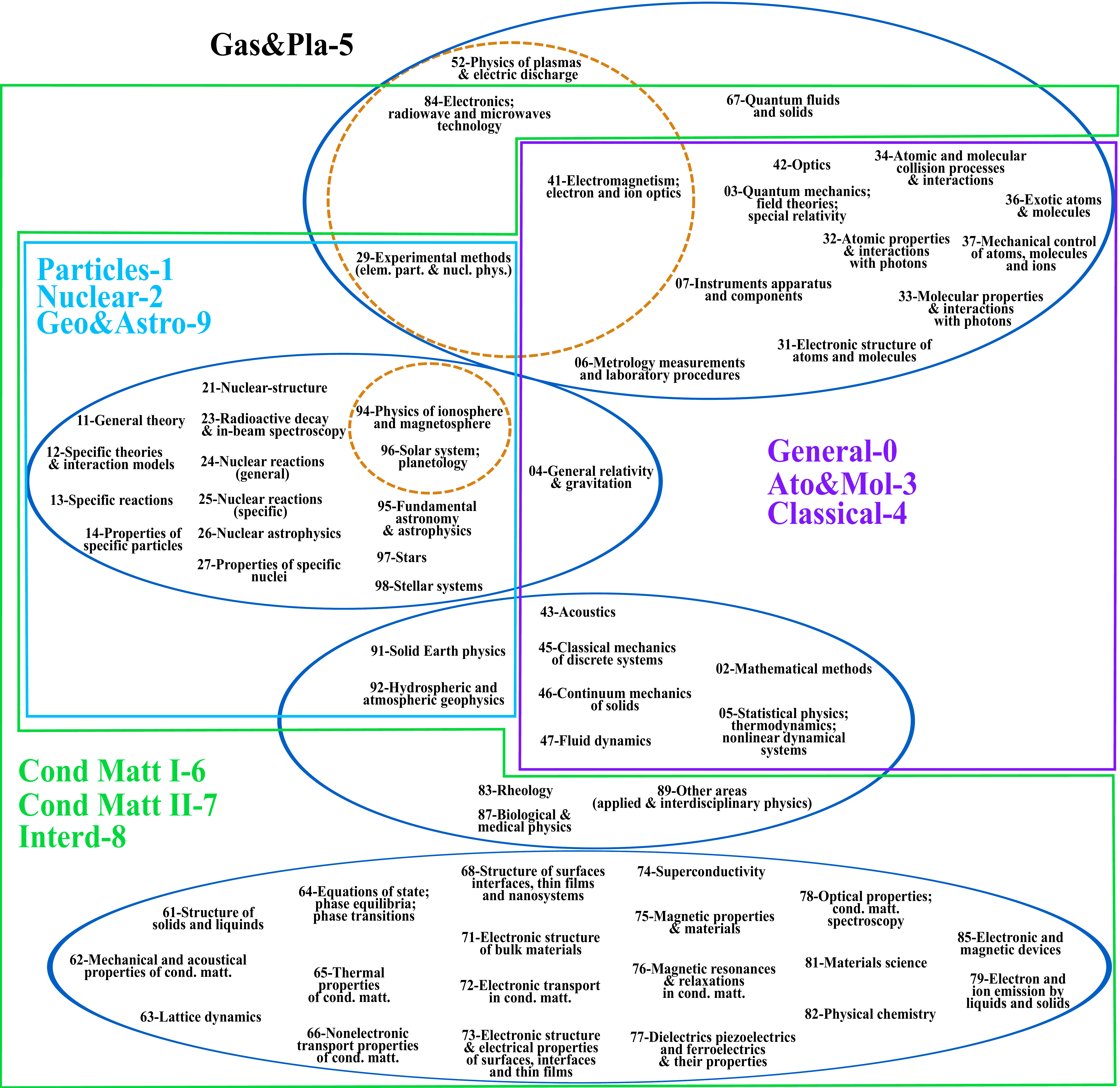} 
  \caption{(Color online) Optimal community structure  of  the layers of the APS Collaboration Network in which each layer represents a collaboration network in a specific area of physics, as described by the second hierarchical level of the PACS code. The four communities found starting from the Infomap-$\widetilde\Theta^{S}_{\alpha,\beta}$ matrix are represented by blue solid-line ovals. In the partition obtained from the Louvain-$\widetilde\Theta^{S}_{\alpha,\beta}$ two sub-communities (ocher dashed ovals) are considered separate communities. These communities form the course-grained partition into the three blocks found at the first hierarchical level of the PACS code (colored solid-line polygons). The nodes displayed in this figure correspond to a subset of $61$ layers that are not isolated in the optimal partition in communities which optimizes the weighted modularity $Q$.}
  \label{fig6}
\end{figure*}

\begin{figure*}
\includegraphics[width=16cm]{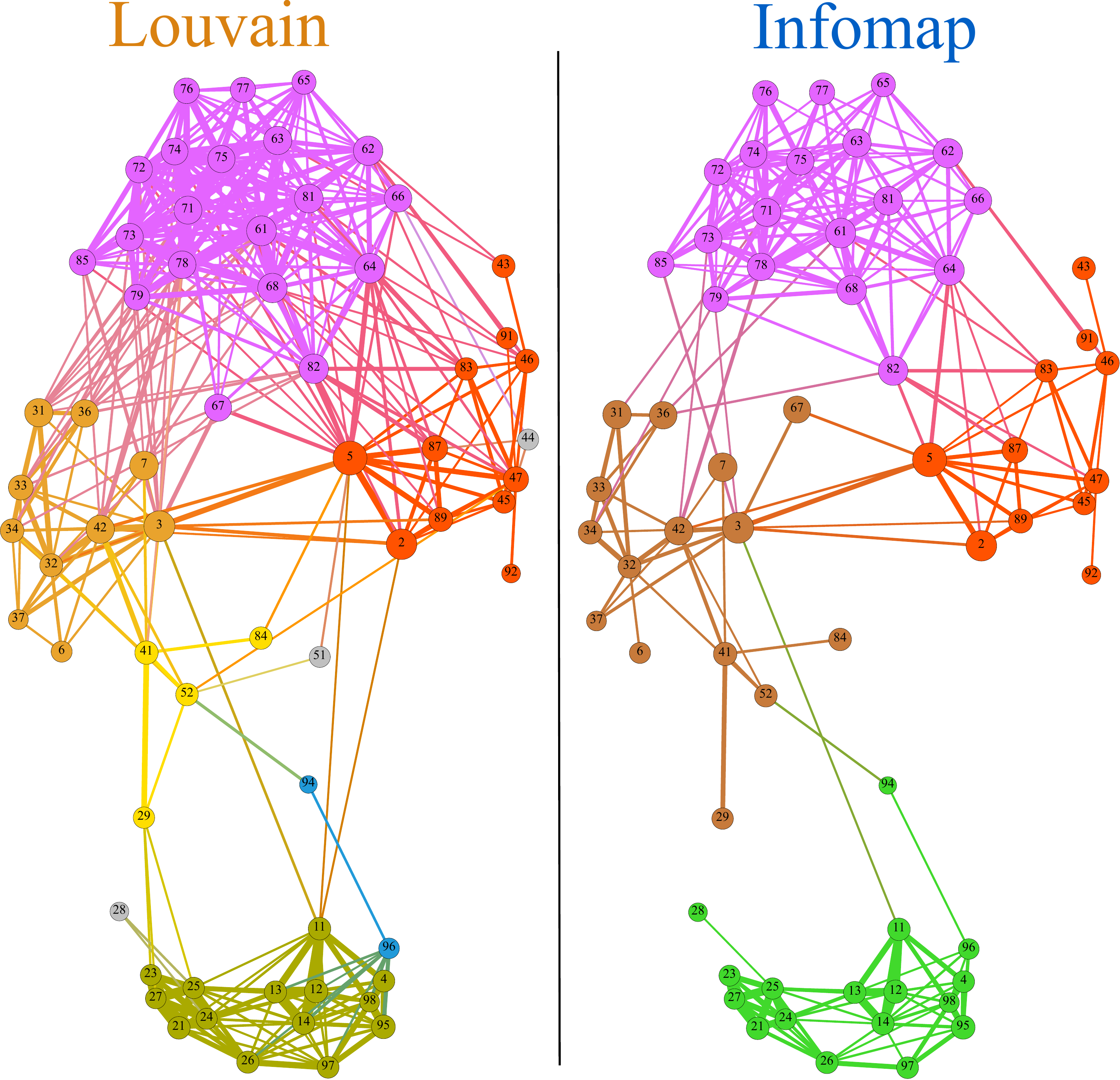} 
  \caption{(Color online) The network between the layers of the APS Collaboration Multiplex Network (with layers corresponding to the PACS code at the second level of the PACS hierarchy) is displayed here for the two cases in which the Louvain-$\widetilde{\Theta}^S$ or the Infomap-$\widetilde{\Theta}^S$ similarity matrix are used. The link weights represent the similarity between the community structure of the two linked layers. The networks are obtained from the $\widetilde{\Theta}^S$ similarity matrix  by filtering out the links below a given threshold value. The threshold is chosen to be the maximal value that ensures that in the filtered network each layer is connected with at least one layer inside its own cluster. The architecture of the networks describes the interplay between the collaboration networks and the organization of knowledge in physics. The community structure revealed by the hierarchical clustering analysis is shown making use of the same color scheme of Figure \ref{fig5}.}
  \label{fig7}
\end{figure*}
In Figure $\ref{fig5}$ we plot the dendrograms resulting from the hierarchical clustering analysis in the case of Louvain-$d$ dissimilarity and Infomap-$d$ dissimilarity. For each dendrogram, the clusters found in the optimal partitions are represented as branches of the same colors. When using the Louvain-$d$ dissimilarity we obtain six clusters plus some isolated layers. When using the Infomap-$d$ dissimilarity we obtain four clusters plus isolated layers. Nevertheless we observe that two of the clusters (the red and the violet clusters) are identically the same in the two partitions. The other two clusters obtained with the Infomap-$d$ dissimilarity are each divided into two clusters when considering the optimal partition using the Louvain-$d$ dissimilarity. In particular the combination of the green-yellow and green-blue  clusters in the Louvain partition is identical to the green cluster of the Infomap partition, while the combination of the orange and the yellow clusters in the Louvain algorithm is identical to the brown cluster of the Infomap partition. 

In Figure $\ref{fig6}$ we give an overview of the blocks hierarchy found. The four clusters found in the Infomap-$d$ optimal partition matrix are represented by solid-line ovals. Dashed ovals split two clusters in two, according to the results obtained from the Louvain-$d$ optimal partition. The block structure at the first level of the PACS hierarchy is shown using solid-line polygons. 
This method allows us to characterize with a bottom-up method how the organization of knowledge in physics is effectively perceived by scientists while shaping their collaboration network.
We observe that while the PACS hierarchy clearly captures main features of the collaboration network, the analysis of the Collaboration Multiplex Network at the second level of the PACS hierarchy clearly suggests a hierarchical organization of these PACS numbers that is not equivalent to the first level of the PACS hierarchy.
Finally we used the information gained by this analysis to construct the network of networks between the layers of the Collaboration Multiplex Network at the second level of the PACS hierarchy.
To this aim we have constructed the weighted network determined by an opportune thresholding of the Louvain-$\widetilde{\Theta}^S$ or Infomap-$\widetilde{\Theta}^S$ similarity matrix (see Figure \ref{fig7}).
The threshold, is here given by the minimum value of the similarity matrix $\widetilde{\Theta}^S$ that ensures that each layer is connected to at least one other layer of its own cluster.
From these networks, it is possible to appreciate that, although the network between the layer of the Collaboration Mutliplex Network is highly interconnected, the clusters found corresponds to layers much more similar between themselves than with other layers outside their own cluster. { Interestingly this visualization shows that the two clusters detected only by the Louvain algorithm, $\left[94,96\right]$ and $\left[29,41,52,84\right]$, contain the nodes that act as bridges
between the yellow-green cluster and the red and the orange clusters. This might explain why the Louvain algorithm identifies them as separate clusters.}

\section{Comparison of the results obtained with $\widetilde\Theta^{S}$ respect to other similarity measures}

In this Section we compare the results obtained from the analysis of the APS Collaboration Multiplex Network using the $\widetilde\Theta^{S}$ indicator with results from other similarity measures commonly used to compare different network partitions \cite{fortunato2010community} and with the $ACTIVS$ Index, an index able to capture the similarity of the layers of a multiplex due to the activity of the nodes. In particular, focusing on the highest level of the PACS hierarchy, we compute the Normalized Mutual Information $NMI$\cite{danon2005comparing}, the Jaccard index  $J$ \cite{jaccard1912distribution}, the Rand index $R$ \cite{rand1971objective,kuncheva2004using} and the $ACTIVIS$ Index for each pair of the $M_1=10$ layers. 
Given two network partitions $X$ and $Y$, the Normalized Mutual Information $NMI$,  is defined as 
\begin{equation}
NMI\left(X,Y\right)=\frac{2\left[H(X)-H(X|Y)\right]}{H(X)+H(Y)},
\end{equation}
where $H(X)=-\sum_xP(x)\log{P(x)}$ is the entropy associated to the distribution $P(x)$ of sizes $x$ of the clusters classified by the partition $X$; $H(Y)$ corresponds to the entropy associated to the distribution $P(y)$ of the sizes $y$ of the clusters in the partition $Y$; $H(X|Y)$ is the conditional entropy associated to the  distribution of the community assignment $X$ conditioned on the distribution of the community assignment $Y$ and  is given by $H(X|Y)=-\sum_{x,y}P(x,y)\log{P(x,y)/P(y)}$, $P(x,y)$ the distribution of the number of nodes having community assignment $x$ in partition $X$ and $y$ in partition $Y$.

The Jaccard index $J$ and the Rand Index $R$, are instead defined as   
\bea
J\left(X,Y\right)&=&\frac{a_{11}}{a_{11}+a_{10}+a_{01}},\nonumber \\
R\left(X,Y\right)&=&\frac{a_{11}+a_{00}}{a_{11}+a_{10}+a_{01}+a_{00}},
\eea
where $a_{11}$ is the number of  pairs of nodes belonging to the same cluster in both partitions $X$ and $Y$, $a_{00}$ is the number of pairs of nodes classified in different clusters in both the $X$ and $Y$ partitions, and $a_{10}$($a_{01}$) is the number of pair of nodes belonging to the same cluster in $X$($Y$) but belonging to different clusters in $Y$($X$).

Finally we define  the Activity Similarity  $ACTIVIS$ Index between the layers $\alpha$ and $\beta$ of a multiplex network, which compares the activity patterns in different layers.
This index is given by 
\bea
ACTIVIS=b_{11}+b_{00},
\eea
where $b_{11}$ are the fraction of nodes active in both layers and $b_{00}$ are the fraction of nodes inactive in both layers.

\begin{figure*}
\includegraphics[width=10cm]{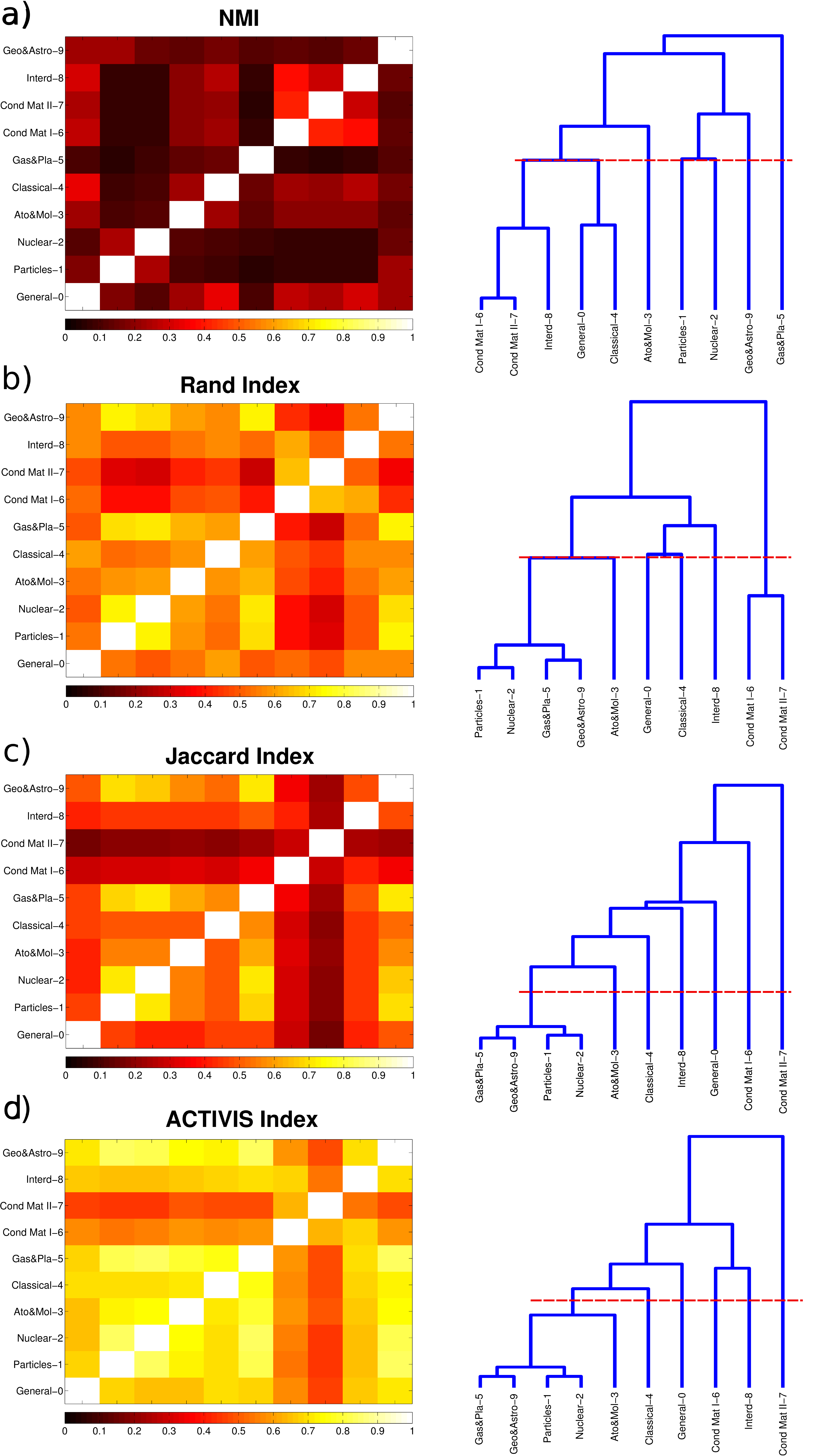} 
  \caption{(Color online) Other similarity measures used to hierarchically cluster the $M_1=10$ layers of the APS Collaboration Multiplex Network at the first level of the PACS hierarchy. The similarity matrices and their respective dendrograms cutted at the partition optimizing the modularity $Q$ (red-dashed line) are shown for the Normalized Mutual Information (Panel a), Rand Index (Panel b), Jacquard Index (Panel c) and for the  $ACTIVIS$ Index (Panel d). Layer partitions are obtained using the Infomap community detection algorithm. None of the optimal partitions corresponds to the one obtained using $\widetilde\Theta^{S}$ to measure similarities.}
  \label{fig8}
\end{figure*}

In Figure \ref{fig8} we show the similarity matrices for the different measures and their respective dendrograms, obtained with the same hierarchical clustering analysis discussed above for the $\widetilde\Theta^{S}$ case.
Here the  layer partitions are obtained using the Infomap algorithm. 
When the modularity $Q$ is optimized, the partition obtained with all these alternative measures are different from the one obtained using the $\widetilde\Theta^{S}$ indicator function.
Moreover the partitions obtained are characterized by having at least $3$ out of $10$ layers in separate clusters, resulting in  significantly less  relevant partitions.
Moreover, by looking at the dendrograms, we can see that none of the other measure is able to give the optimal partition obtained with $\widetilde\Theta^{S}$ even by applying an arbitrary cut to the respective dendrogram.

These results show clearly that  the proposed indicator function  $\widetilde\Theta^{S}$ based on information theory, is not equivalent to previously defined similarity measures between partitions. Moreover the method is not affected significantly by the choice we made for treating inactive nodes or nodes belonging to connected components of two nodes.
Although it might be a challenging technical problem to assess which of the similarity measures proposed so far is the best, the similarity measure $\widetilde\Theta^{S}$ seems to be more relevant of other similarity measures used in the literature when applied to the APS Collaboration Multiplex Networks. In fact the partition obtained by using the similarity measure  $\widetilde\Theta^{S}$  reflect much more closely the general perception of the organization of collaborations in the physics community. 
 
\section{Conclusions}

Characterizing the mesoscopic structure of multiplex networks is crucial to characterize large network datasets where the nodes are connected by different types of interactions. Such multilayer networks are ubiquitous, and systems as different as social networks, transportation networks or cellular and brain networks require a multilayer description.
Here, by using information theory tools, we have defined an indicator function $\widetilde\Theta^{S}$ able to measure the mesoscopic similarities between the layers of a multiplex network. This indicator can be used to quantitatively compare the layers of a multiplex network with respect to the mesoscopic structure induced by any feature depending on the layer architecture. In particular here we have focused on the case in which the feature of the nodes is their community assignment. We have shown that $\widetilde\Theta^{S}$ can reveal the network between the layers of a multiplex and we have applied this method to the Collaboration Multiplex Network at the two levels of the PACS hierarchy, obtaining a bottom-up approach to identify how the organization of knowledge in physics is reflected in the structure of collaboration networks.


\begin{thebibliography}{38}
\expandafter\ifx\csname natexlab\endcsname\relax\def\natexlab#1{#1}\fi
\expandafter\ifx\csname bibnamefont\endcsname\relax
  \def\bibnamefont#1{#1}\fi
\expandafter\ifx\csname bibfnamefont\endcsname\relax
  \def\bibfnamefont#1{#1}\fi
\expandafter\ifx\csname citenamefont\endcsname\relax
  \def\citenamefont#1{#1}\fi
\expandafter\ifx\csname url\endcsname\relax
  \def\url#1{\texttt{#1}}\fi
\expandafter\ifx\csname urlprefix\endcsname\relax\def\urlprefix{URL }\fi
\providecommand{\bibinfo}[2]{#2}
\providecommand{\eprint}[2][]{\url{#2}}

\bibitem[{\citenamefont{Boccaletti et~al.}(2014)\citenamefont{Boccaletti,
  Bianconi, Criado, Del~Genio, G{\'o}mez-Garde{\~n}es, Romance, Sendina-Nadal,
  Wang, and Zanin}}]{boccaletti2014structure}
\bibinfo{author}{\bibfnamefont{S.}~\bibnamefont{Boccaletti}},
  \bibinfo{author}{\bibfnamefont{G.}~\bibnamefont{Bianconi}},
  \bibinfo{author}{\bibfnamefont{R.}~\bibnamefont{Criado}},
  \bibinfo{author}{\bibfnamefont{C.}~\bibnamefont{Del~Genio}},
  \bibinfo{author}{\bibfnamefont{J.}~\bibnamefont{G{\'o}mez-Garde{\~n}es}},
  \bibinfo{author}{\bibfnamefont{M.}~\bibnamefont{Romance}},
  \bibinfo{author}{\bibfnamefont{I.}~\bibnamefont{Sendina-Nadal}},
  \bibinfo{author}{\bibfnamefont{Z.}~\bibnamefont{Wang}}, \bibnamefont{and}
  \bibinfo{author}{\bibfnamefont{M.}~\bibnamefont{Zanin}},
  \bibinfo{journal}{Physics Reports} \textbf{\bibinfo{volume}{544}},
  \bibinfo{pages}{1} (\bibinfo{year}{2014}).

\bibitem[{\citenamefont{Kivel{\"a} et~al.}(2014)\citenamefont{Kivel{\"a},
  Arenas, Barthelemy, Gleeson, Moreno, and Porter}}]{kivela2014multilayer}
\bibinfo{author}{\bibfnamefont{M.}~\bibnamefont{Kivel{\"a}}},
  \bibinfo{author}{\bibfnamefont{A.}~\bibnamefont{Arenas}},
  \bibinfo{author}{\bibfnamefont{M.}~\bibnamefont{Barthelemy}},
  \bibinfo{author}{\bibfnamefont{J.~P.} \bibnamefont{Gleeson}},
  \bibinfo{author}{\bibfnamefont{Y.}~\bibnamefont{Moreno}}, \bibnamefont{and}
  \bibinfo{author}{\bibfnamefont{M.~A.} \bibnamefont{Porter}},
  \bibinfo{journal}{Journal of Complex Networks} \textbf{\bibinfo{volume}{2}},
  \bibinfo{pages}{203} (\bibinfo{year}{2014}).

\bibitem[{\citenamefont{Szell et~al.}(2010)\citenamefont{Szell, Lambiotte, and
  Thurner}}]{szell2010multirelational}
\bibinfo{author}{\bibfnamefont{M.}~\bibnamefont{Szell}},
  \bibinfo{author}{\bibfnamefont{R.}~\bibnamefont{Lambiotte}},
  \bibnamefont{and} \bibinfo{author}{\bibfnamefont{S.}~\bibnamefont{Thurner}},
  \bibinfo{journal}{Proceedings of the National Academy of Sciences}
  \textbf{\bibinfo{volume}{107}}, \bibinfo{pages}{13636}
  (\bibinfo{year}{2010}).

\bibitem[{\citenamefont{Cardillo et~al.}(2012)\citenamefont{Cardillo, Zanin,
  G{\'o}mez-Garde{\~n}es, Romance, del Amo, and
  Boccaletti}}]{cardillo2012modeling}
\bibinfo{author}{\bibfnamefont{A.}~\bibnamefont{Cardillo}},
  \bibinfo{author}{\bibfnamefont{M.}~\bibnamefont{Zanin}},
  \bibinfo{author}{\bibfnamefont{J.}~\bibnamefont{G{\'o}mez-Garde{\~n}es}},
  \bibinfo{author}{\bibfnamefont{M.}~\bibnamefont{Romance}},
  \bibinfo{author}{\bibfnamefont{A.~J.~G.} \bibnamefont{del Amo}},
  \bibnamefont{and}
  \bibinfo{author}{\bibfnamefont{S.}~\bibnamefont{Boccaletti}},
  \bibinfo{journal}{arXiv preprint arXiv:1211.6839}  (\bibinfo{year}{2012}).

\bibitem[{\citenamefont{Menichetti et~al.}(2014)\citenamefont{Menichetti,
  Remondini, Panzarasa, Mondrag{\'o}n, and Bianconi}}]{menichetti2014weighted}
\bibinfo{author}{\bibfnamefont{G.}~\bibnamefont{Menichetti}},
  \bibinfo{author}{\bibfnamefont{D.}~\bibnamefont{Remondini}},
  \bibinfo{author}{\bibfnamefont{P.}~\bibnamefont{Panzarasa}},
  \bibinfo{author}{\bibfnamefont{R.~J.} \bibnamefont{Mondrag{\'o}n}},
  \bibnamefont{and} \bibinfo{author}{\bibfnamefont{G.}~\bibnamefont{Bianconi}},
  \bibinfo{journal}{PloS one} \textbf{\bibinfo{volume}{9}},
  \bibinfo{pages}{e97857} (\bibinfo{year}{2014}).

\bibitem[{\citenamefont{Nicosia and Latora}(2014)}]{nicosia2014measuring}
\bibinfo{author}{\bibfnamefont{V.}~\bibnamefont{Nicosia}} \bibnamefont{and}
  \bibinfo{author}{\bibfnamefont{V.}~\bibnamefont{Latora}},
  \bibinfo{journal}{arXiv preprint arXiv:1403.1546}  (\bibinfo{year}{2014}).

\bibitem[{\citenamefont{Bullmore and Sporns}(2009)}]{bullmore2009complex}
\bibinfo{author}{\bibfnamefont{E.}~\bibnamefont{Bullmore}} \bibnamefont{and}
  \bibinfo{author}{\bibfnamefont{O.}~\bibnamefont{Sporns}},
  \bibinfo{journal}{Nature Reviews Neuroscience} \textbf{\bibinfo{volume}{10}},
  \bibinfo{pages}{186} (\bibinfo{year}{2009}).

\bibitem[{\citenamefont{Mucha et~al.}(2010)\citenamefont{Mucha, Richardson,
  Macon, Porter, and Onnela}}]{mucha2010community}
\bibinfo{author}{\bibfnamefont{P.~J.} \bibnamefont{Mucha}},
  \bibinfo{author}{\bibfnamefont{T.}~\bibnamefont{Richardson}},
  \bibinfo{author}{\bibfnamefont{K.}~\bibnamefont{Macon}},
  \bibinfo{author}{\bibfnamefont{M.~A.} \bibnamefont{Porter}},
  \bibnamefont{and} \bibinfo{author}{\bibfnamefont{J.-P.}
  \bibnamefont{Onnela}}, \bibinfo{journal}{science}
  \textbf{\bibinfo{volume}{328}}, \bibinfo{pages}{876} (\bibinfo{year}{2010}).

\bibitem[{\citenamefont{Battiston et~al.}(2014)\citenamefont{Battiston,
  Nicosia, and Latora}}]{battiston2014structural}
\bibinfo{author}{\bibfnamefont{F.}~\bibnamefont{Battiston}},
  \bibinfo{author}{\bibfnamefont{V.}~\bibnamefont{Nicosia}}, \bibnamefont{and}
  \bibinfo{author}{\bibfnamefont{V.}~\bibnamefont{Latora}},
  \bibinfo{journal}{Physical Review E} \textbf{\bibinfo{volume}{89}},
  \bibinfo{pages}{032804} (\bibinfo{year}{2014}).


\bibitem[{\citenamefont{De~Domenico
  et~al.}(2015{\natexlab{a}})\citenamefont{De~Domenico, Nicosia, Arenas, and
  Latora}}]{de2015structural}
\bibinfo{author}{\bibfnamefont{M.}~\bibnamefont{De~Domenico}},
  \bibinfo{author}{\bibfnamefont{V.}~\bibnamefont{Nicosia}},
  \bibinfo{author}{\bibfnamefont{A.}~\bibnamefont{Arenas}}, \bibnamefont{and}
  \bibinfo{author}{\bibfnamefont{V.}~\bibnamefont{Latora}},
  \bibinfo{journal}{Nature communications} \textbf{\bibinfo{volume}{6}}
  (\bibinfo{year}{2015}{\natexlab{a}}).


\bibitem[{\citenamefont{Halu et~al.}(2013)\citenamefont{Halu, Mondrag{\'o}n,
  Panzarasa, and Bianconi}}]{halu2013multiplex}
\bibinfo{author}{\bibfnamefont{A.}~\bibnamefont{Halu}},
  \bibinfo{author}{\bibfnamefont{R.~J.} \bibnamefont{Mondrag{\'o}n}},
  \bibinfo{author}{\bibfnamefont{P.}~\bibnamefont{Panzarasa}},
  \bibnamefont{and} \bibinfo{author}{\bibfnamefont{G.}~\bibnamefont{Bianconi}},
  \bibinfo{journal}{PloS one} \textbf{\bibinfo{volume}{8}},
  \bibinfo{pages}{e78293} (\bibinfo{year}{2013}).

\bibitem[{\citenamefont{Sol{\'a} et~al.}(2013)\citenamefont{Sol{\'a}, Romance,
  Criado, Flores, del Amo, and Boccaletti}}]{sola2013eigenvector}
\bibinfo{author}{\bibfnamefont{L.}~\bibnamefont{Sol{\'a}}},
  \bibinfo{author}{\bibfnamefont{M.}~\bibnamefont{Romance}},
  \bibinfo{author}{\bibfnamefont{R.}~\bibnamefont{Criado}},
  \bibinfo{author}{\bibfnamefont{J.}~\bibnamefont{Flores}},
  \bibinfo{author}{\bibfnamefont{A.~G.} \bibnamefont{del Amo}},
  \bibnamefont{and}
  \bibinfo{author}{\bibfnamefont{S.}~\bibnamefont{Boccaletti}},
  \bibinfo{journal}{Chaos: An Interdisciplinary Journal of Nonlinear Science}
  \textbf{\bibinfo{volume}{23}}, \bibinfo{pages}{033131}
  (\bibinfo{year}{2013}).


\bibitem[{\citenamefont{De~Domenico
  et~al.}(2015{\natexlab{b}})\citenamefont{De~Domenico, Sol{\'e}-Ribalta,
  Omodei, G{\'o}mez, and Arenas}}]{de2015ranking}
\bibinfo{author}{\bibfnamefont{M.}~\bibnamefont{De~Domenico}},
  \bibinfo{author}{\bibfnamefont{A.}~\bibnamefont{Sol{\'e}-Ribalta}},
  \bibinfo{author}{\bibfnamefont{E.}~\bibnamefont{Omodei}},
  \bibinfo{author}{\bibfnamefont{S.}~\bibnamefont{G{\'o}mez}},
  \bibnamefont{and} \bibinfo{author}{\bibfnamefont{A.}~\bibnamefont{Arenas}},
  \bibinfo{journal}{Nature communications} \textbf{\bibinfo{volume}{6}}
  (\bibinfo{year}{2015}{\natexlab{b}}).

\bibitem[{\citenamefont{Bianconi}(2013)}]{bianconi2013statistical}
\bibinfo{author}{\bibfnamefont{G.}~\bibnamefont{Bianconi}},
  \bibinfo{journal}{Physical Review E} \textbf{\bibinfo{volume}{87}},
  \bibinfo{pages}{062806} (\bibinfo{year}{2013}).

\bibitem[{\citenamefont{Min et~al.}(2014)\citenamefont{Min, Do~Yi, Lee, and
  Goh}}]{min2014network}
\bibinfo{author}{\bibfnamefont{B.}~\bibnamefont{Min}},
  \bibinfo{author}{\bibfnamefont{S.}~\bibnamefont{Do~Yi}},
  \bibinfo{author}{\bibfnamefont{K.-M.} \bibnamefont{Lee}}, \bibnamefont{and}
  \bibinfo{author}{\bibfnamefont{K.-I.} \bibnamefont{Goh}},
  \bibinfo{journal}{Physical Review E} \textbf{\bibinfo{volume}{89}},
  \bibinfo{pages}{042811} (\bibinfo{year}{2014}).

\bibitem[{\citenamefont{Cellai and Bianconi}(2015)}]{cellai2015}
\bibinfo{author}{\bibfnamefont{D.}~\bibnamefont{Cellai}} \bibnamefont{and}
  \bibinfo{author}{\bibfnamefont{G.}~\bibnamefont{Bianconi}},
  \bibinfo{journal}{arXiv preprint arXiv:1505.01220}  (\bibinfo{year}{2015}).

\bibitem[{\citenamefont{Fortunato}(2010)}]{fortunato2010community}
\bibinfo{author}{\bibfnamefont{S.}~\bibnamefont{Fortunato}},
  \bibinfo{journal}{Physics Reports} \textbf{\bibinfo{volume}{486}},
  \bibinfo{pages}{75} (\bibinfo{year}{2010}).

\bibitem[{\citenamefont{Bianconi et~al.}(2014)\citenamefont{Bianconi, Darst,
  Iacovacci, and Fortunato}}]{bianconi2014triadic}
\bibinfo{author}{\bibfnamefont{G.}~\bibnamefont{Bianconi}},
  \bibinfo{author}{\bibfnamefont{R.~K.} \bibnamefont{Darst}},
  \bibinfo{author}{\bibfnamefont{J.}~\bibnamefont{Iacovacci}},
  \bibnamefont{and}
  \bibinfo{author}{\bibfnamefont{S.}~\bibnamefont{Fortunato}},
  \bibinfo{journal}{Physical Review E} \textbf{\bibinfo{volume}{90}},
  \bibinfo{pages}{042806} (\bibinfo{year}{2014}).

\bibitem[{\citenamefont{De~Domenico
  et~al.}(2015{\natexlab{c}})\citenamefont{De~Domenico, Lancichinetti, Arenas,
  and Rosvall}}]{de2014identifying}
\bibinfo{author}{\bibfnamefont{M.}~\bibnamefont{De~Domenico}},
  \bibinfo{author}{\bibfnamefont{A.}~\bibnamefont{Lancichinetti}},
  \bibinfo{author}{\bibfnamefont{A.}~\bibnamefont{Arenas}}, \bibnamefont{and}
  \bibinfo{author}{\bibfnamefont{M.}~\bibnamefont{Rosvall}},
  \bibinfo{journal}{Phys. Rev. X} \textbf{\bibinfo{volume}{5}}
  (\bibinfo{year}{2015}{\natexlab{c}}).


\bibitem[{\citenamefont{Valles-Catala et~al.}(2014)\citenamefont{Valles-Catala,
  Massucci, Guimera, and Sales-Pardo}}]{valles2014multilayer}
\bibinfo{author}{\bibfnamefont{T.}~\bibnamefont{Valles-Catala}},
  \bibinfo{author}{\bibfnamefont{F.~A.} \bibnamefont{Massucci}},
  \bibinfo{author}{\bibfnamefont{R.}~\bibnamefont{Guimera}}, \bibnamefont{and}
  \bibinfo{author}{\bibfnamefont{M.}~\bibnamefont{Sales-Pardo}},
  \bibinfo{journal}{arXiv preprint arXiv:1411.1098}  (\bibinfo{year}{2014}).

\bibitem[{\citenamefont{Manlio De~Domenico}(2014)}]{de2014multilayer}
\bibinfo{author}{\bibfnamefont{A.~A.} \bibnamefont{Manlio De~Domenico},
  \bibfnamefont{Mason A.~Porter}}, \bibinfo{journal}{Journal of Complex
  Networks, doi: 10.1093/comnet/cnu038 (2014)}  (\bibinfo{year}{2014}).

\bibitem[{\citenamefont{Battiston et~al.}(2015)\citenamefont{Battiston,
  Iacovacci, Nicosia, Bianconi, and Latora}}]{battiston2015emergence}
\bibinfo{author}{\bibfnamefont{F.}~\bibnamefont{Battiston}},
  \bibinfo{author}{\bibfnamefont{J.}~\bibnamefont{Iacovacci}},
  \bibinfo{author}{\bibfnamefont{V.}~\bibnamefont{Nicosia}},
  \bibinfo{author}{\bibfnamefont{G.}~\bibnamefont{Bianconi}}, \bibnamefont{and}
  \bibinfo{author}{\bibfnamefont{V.}~\bibnamefont{Latora}},
  \bibinfo{journal}{arXiv preprint arXiv:1506.01280}  (\bibinfo{year}{2015}).


\bibitem[{\citenamefont{Bianconi}(2008)}]{bianconi2008entropy}
\bibinfo{author}{\bibfnamefont{G.}~\bibnamefont{Bianconi}},
  \bibinfo{journal}{EPL (Europhysics Letters)} \textbf{\bibinfo{volume}{81}},
  \bibinfo{pages}{28005} (\bibinfo{year}{2008}).

\bibitem[{\citenamefont{Bianconi}(2009)}]{bianconi2009entropy}
\bibinfo{author}{\bibfnamefont{G.}~\bibnamefont{Bianconi}},
  \bibinfo{journal}{Physical Review E} \textbf{\bibinfo{volume}{79}},
  \bibinfo{pages}{036114} (\bibinfo{year}{2009}).

\bibitem[{\citenamefont{Peixoto}(2012)}]{peixoto2012entropy}
\bibinfo{author}{\bibfnamefont{T.~P.} \bibnamefont{Peixoto}},
  \bibinfo{journal}{Physical Review E} \textbf{\bibinfo{volume}{85}},
  \bibinfo{pages}{056122} (\bibinfo{year}{2012}).

\bibitem[{\citenamefont{Bianconi et~al.}(2009)\citenamefont{Bianconi, Pin, and
  Marsili}}]{bianconi2009assessing}
\bibinfo{author}{\bibfnamefont{G.}~\bibnamefont{Bianconi}},
  \bibinfo{author}{\bibfnamefont{P.}~\bibnamefont{Pin}}, \bibnamefont{and}
  \bibinfo{author}{\bibfnamefont{M.}~\bibnamefont{Marsili}},
  \bibinfo{journal}{Proceedings of the National Academy of Sciences}
  \textbf{\bibinfo{volume}{106}}, \bibinfo{pages}{11433}
  (\bibinfo{year}{2009}).

\bibitem[{dat()}]{dataset}
\emph{\bibinfo{title}{Aps data sets for research {@ONLINE}}},
  \urlprefix\url{http://journals.aps.org/datasets}.

\bibitem[{\citenamefont{Redner}(1998)}]{redner98}
\bibinfo{author}{\bibfnamefont{S.}~\bibnamefont{Redner}},
  \bibinfo{journal}{Eur. Phys. J. B} \textbf{\bibinfo{volume}{4}},
  \bibinfo{pages}{131} (\bibinfo{year}{1998}).

\bibitem[{\citenamefont{Newman}(2001{\natexlab{a}})}]{newman2001PNAS}
\bibinfo{author}{\bibfnamefont{M.~E.~J.} \bibnamefont{Newman}},
  \bibinfo{journal}{Proceedings of the National Academy of Science}
  \textbf{\bibinfo{volume}{4}}, \bibinfo{pages}{404}
  (\bibinfo{year}{2001}{\natexlab{a}}).

\bibitem[{\citenamefont{Newman}(2001{\natexlab{b}})}]{newman2001PRE}
\bibinfo{author}{\bibfnamefont{M.~E.~J.} \bibnamefont{Newman}},
  \bibinfo{journal}{Physical Review E} \textbf{\bibinfo{volume}{64}},
  \bibinfo{pages}{016132} (\bibinfo{year}{2001}{\natexlab{b}}).

\bibitem[{\citenamefont{Arenas et~al.}(2004)\citenamefont{Arenas, Danon,
  Diaz-Guilera, Gleiser, and Guimera}}]{arenas2004}
\bibinfo{author}{\bibfnamefont{A.}~\bibnamefont{Arenas}},
  \bibinfo{author}{\bibfnamefont{L.}~\bibnamefont{Danon}},
  \bibinfo{author}{\bibfnamefont{A.}~\bibnamefont{Diaz-Guilera}},
  \bibinfo{author}{\bibfnamefont{P.~M.} \bibnamefont{Gleiser}},
  \bibnamefont{and} \bibinfo{author}{\bibfnamefont{R.}~\bibnamefont{Guimera}},
  \bibinfo{journal}{The European Physical Journal B}
  \textbf{\bibinfo{volume}{38}}, \bibinfo{pages}{373} (\bibinfo{year}{2004}).

\bibitem[{\citenamefont{Lee et~al.}(2010)\citenamefont{Lee, Goh, Kahng, and
  Kim}}]{lee2010}
\bibinfo{author}{\bibfnamefont{D.}~\bibnamefont{Lee}},
  \bibinfo{author}{\bibfnamefont{K.-I.} \bibnamefont{Goh}},
  \bibinfo{author}{\bibfnamefont{B.}~\bibnamefont{Kahng}}, \bibnamefont{and}
  \bibinfo{author}{\bibfnamefont{D.}~\bibnamefont{Kim}},
  \bibinfo{journal}{Physical Review E} \textbf{\bibinfo{volume}{82}},
  \bibinfo{pages}{026112} (\bibinfo{year}{2010}).



\bibitem[{pac()}]{pacs}
\emph{\bibinfo{title}{Pacs 2010 regular edition - american institute of physics
  {@ONLINE}}},
  \urlprefix\url{http://www.aip.org/publishing/pacs/pacs-2010-regular-edition}.

\bibitem[{\citenamefont{Girvan and Newman}(2002)}]{girvan2002community}
\bibinfo{author}{\bibfnamefont{M.}~\bibnamefont{Girvan}} \bibnamefont{and}
  \bibinfo{author}{\bibfnamefont{M.~E.} \bibnamefont{Newman}},
  \bibinfo{journal}{Proceedings of the National Academy of Sciences}
  \textbf{\bibinfo{volume}{99}}, \bibinfo{pages}{7821} (\bibinfo{year}{2002}).

\bibitem[{\citenamefont{Lancichinetti et~al.}(2008)\citenamefont{Lancichinetti,
  Fortunato, and Radicchi}}]{lancichinetti2008benchmark}
\bibinfo{author}{\bibfnamefont{A.}~\bibnamefont{Lancichinetti}},
  \bibinfo{author}{\bibfnamefont{S.}~\bibnamefont{Fortunato}},
  \bibnamefont{and} \bibinfo{author}{\bibfnamefont{F.}~\bibnamefont{Radicchi}},
  \bibinfo{journal}{Physical review E} \textbf{\bibinfo{volume}{78}},
  \bibinfo{pages}{046110} (\bibinfo{year}{2008}).

\bibitem[{\citenamefont{Rosvall and Bergstrom}(2007)}]{rosvall2007information}
\bibinfo{author}{\bibfnamefont{M.}~\bibnamefont{Rosvall}} \bibnamefont{and}
  \bibinfo{author}{\bibfnamefont{C.~T.} \bibnamefont{Bergstrom}},
  \bibinfo{journal}{Proceedings of the National Academy of Sciences}
  \textbf{\bibinfo{volume}{104}}, \bibinfo{pages}{7327} (\bibinfo{year}{2007}).

\bibitem[{\citenamefont{Blondel et~al.}(2008)\citenamefont{Blondel, Guillaume,
  Lambiotte, and Lefebvre}}]{blondel2008fast}
\bibinfo{author}{\bibfnamefont{V.~D.} \bibnamefont{Blondel}},
  \bibinfo{author}{\bibfnamefont{J.-L.} \bibnamefont{Guillaume}},
  \bibinfo{author}{\bibfnamefont{R.}~\bibnamefont{Lambiotte}},
  \bibnamefont{and} \bibinfo{author}{\bibfnamefont{E.}~\bibnamefont{Lefebvre}},
  \bibinfo{journal}{Journal of Statistical Mechanics: Theory and Experiment}
  \textbf{\bibinfo{volume}{2008}}, \bibinfo{pages}{P10008}
  (\bibinfo{year}{2008}).

\bibitem[{\citenamefont{Sokal and Michener}(1958)}]{clust1}
\bibinfo{author}{\bibfnamefont{R.}~\bibnamefont{Sokal}} \bibnamefont{and}
  \bibinfo{author}{\bibfnamefont{C.}~\bibnamefont{Michener}},
  \bibinfo{journal}{University of Kansas Science Bulletin}
  \textbf{\bibinfo{volume}{38}}, \bibinfo{pages}{1409} (\bibinfo{year}{1958}).

\bibitem[{\citenamefont{Sokal and Rohlf}(1962)}]{clust2}
\bibinfo{author}{\bibfnamefont{R.}~\bibnamefont{Sokal}} \bibnamefont{and}
  \bibinfo{author}{\bibfnamefont{F.~J.} \bibnamefont{Rohlf}},
  \bibinfo{journal}{Taxon} \textbf{\bibinfo{volume}{11}}, \bibinfo{pages}{33}
  (\bibinfo{year}{1962}).

\bibitem[{\citenamefont{Ying et~al.}(2005)\citenamefont{Ying, Karypis, and
  Fayyad}}]{clust3}
\bibinfo{author}{\bibfnamefont{Z.}~\bibnamefont{Ying}},
  \bibinfo{author}{\bibfnamefont{G.}~\bibnamefont{Karypis}}, \bibnamefont{and}
  \bibinfo{author}{\bibfnamefont{U.}~\bibnamefont{Fayyad}},
  \bibinfo{journal}{Data mining and knowledge discovery}
  \textbf{\bibinfo{volume}{10}}, \bibinfo{pages}{141} (\bibinfo{year}{2005}).

\bibitem[{\citenamefont{Newman}(2006)}]{newman2006modularity}
\bibinfo{author}{\bibfnamefont{M.~E.} \bibnamefont{Newman}},
  \bibinfo{journal}{Proceedings of the National Academy of Sciences}
  \textbf{\bibinfo{volume}{103}}, \bibinfo{pages}{8577} (\bibinfo{year}{2006}).


\bibitem[{\citenamefont{Jaccard}(1912)}]{jaccard1912distribution}
\bibinfo{author}{\bibfnamefont{P.}~\bibnamefont{Jaccard}},
  \bibinfo{journal}{New phytologist} \textbf{\bibinfo{volume}{11}},
  \bibinfo{pages}{37} (\bibinfo{year}{1912}).

\bibitem[{\citenamefont{Rand}(1971)}]{rand1971objective}
\bibinfo{author}{\bibfnamefont{W.~M.} \bibnamefont{Rand}},
  \bibinfo{journal}{Journal of the American Statistical association}
  \textbf{\bibinfo{volume}{66}}, \bibinfo{pages}{846} (\bibinfo{year}{1971}).

\bibitem[{\citenamefont{Kuncheva et~al.}(2004)\citenamefont{Kuncheva,
  Hadjitodorov et~al.}}]{kuncheva2004using}
\bibinfo{author}{\bibfnamefont{L.}~\bibnamefont{Kuncheva}},
  \bibinfo{author}{\bibfnamefont{S.~T.} \bibnamefont{Hadjitodorov}},
  \bibnamefont{et~al.}, in \emph{\bibinfo{booktitle}{Systems, man and
  cybernetics, 2004 IEEE international conference on}}
  (\bibinfo{organization}{IEEE}, \bibinfo{year}{2004}),
  vol.~\bibinfo{volume}{2}, pp. \bibinfo{pages}{1214--1219}.

\bibitem[{\citenamefont{Danon et~al.}(2005)\citenamefont{Danon, Diaz-Guilera,
  Duch, and Arenas}}]{danon2005comparing}
\bibinfo{author}{\bibfnamefont{L.}~\bibnamefont{Danon}},
  \bibinfo{author}{\bibfnamefont{A.}~\bibnamefont{Diaz-Guilera}},
  \bibinfo{author}{\bibfnamefont{J.}~\bibnamefont{Duch}}, \bibnamefont{and}
  \bibinfo{author}{\bibfnamefont{A.}~\bibnamefont{Arenas}},
  \bibinfo{journal}{Journal of Statistical Mechanics: Theory and Experiment}
  \textbf{\bibinfo{volume}{2005}}, \bibinfo{pages}{P09008}
  (\bibinfo{year}{2005}).

\end{thebibliography}
\end{document}